\documentclass[%
 reprint,
 amsmath,amssymb,
 aps,superscriptaddress
]{revtex4-2}

\usepackage[english]{babel}
  \usepackage{cancel}
\usepackage{graphicx}
\usepackage{dcolumn}
\usepackage{bm}
\usepackage{hyperref}
\usepackage{xcolor}

\renewcommand{\vec}[1]{\boldsymbol{#1}}
\renewcommand{\imath}{i}
\newcommand*{\norme}[1]{\left\lVert\mathrm{#1}\right\rVert}

\definecolor{vert2}{RGB}{30, 120, 5}

\def\ea{\textit{et al.}}

\def\Id{\mathbf{I}} 
\def\G{\mathbf{G}_0^\prime} 
\def\H0{\mathcal{H}_0^{(1)}}
\def\Hpp{\mathbf{G_0^{\prime \prime}}} 

\def\Ktf{\mathbf{K}} 
\def\Kf{{\mathbf{K}}^{(1)}} 
\def\Kdf{{\mathbf{K}}^{(2)}} 
\def\Ktrf{{\mathbf{K}}^{(3)}} 
 
\def\Knf{{\mathbf{K}}^{(n)}} 
\def\Kc{\mathbf{K}_{\mathbf{c}}} 
\def\Kpmult{\mathbf{K}_{\mathbf{m \backslash r}}} 
\def\Kpn{\mathbf{K}_{\mathbf{m \backslash r}}^{(n)}} 
\def\Kfilt{\mathbf{K}_{\mathbf{f}}}

\def\Kmult{\mathbf{K}_{\mathbf{m}}} 
 
\def\Krec{\mathbf{K}_{\mathbf{r}}} 
 
\def\Ts{\mathcal{T}_s}
\def\T{\mathcal{T}_0}
\def\En{\mathbf{E}_{\mathbf{l}}}

\def\Rrr{\mathbf{R}} 
\def\Rz{\mathbf{\overline{R}}} 
\def\Rc{\mathbf{\overline{R}}_{\mathbf{c}}}
\def\Rf{\mathbf{\overline{R}}^{(1)}}

\def\Rfilt{\mathbf{\overline{R}}_{\mathbf{f}}} 
\def\Rfiltf{\mathbf{{R}}_{\mathbf{f}}} 

\def\Rpmult{\mathbf{\overline{R}}_{\mathbf{m \backslash r}}} 
\def\Rmult{\mathbf{\overline{R}}_{\mathbf{m}}}

\def\rin{\mathbf{r}_{\mathbf{in}}} 
\def\rout{\mathbf{r}_{\mathbf{out}}} 

\def\Rl{\mathbf{{F}_l}} 
\def\Rll{\mathbf{^\perp { F}_l}} 

\def\ui{\vec{u_i}}
\def\uj{\vec{u_j}}

\def\ri{\vec{r_1}}
\def\rj{\vec{r_2}}

 \newcommand{\dd}{
      \mathop{}\mathopen{}\mathrm{d}
      }

\def\du{\Delta u}

\begin{document}

\preprint{APS/123-QED}

\title{Weight of single and recurrent scattering \\in the reflection matrix of complex media}
\author{Cécile BRÜTT}
\affiliation{Institut Langevin, ESPCI Paris, CNRS, PSL University, 1 rue Jussieu, 75005 Paris, France}%
\affiliation{Safran Tech, Digital Sciences and Technologies Department, Rue des Jeunes Bois, Châteaufort, 78114 Magny-Les-Hameaux, France}%
\author{Alexandre AUBRY}
\affiliation{Institut Langevin, ESPCI Paris, CNRS, PSL University, 1 rue Jussieu, 75005 Paris, France}%
\author{Benoît GÉRARDIN}%
\affiliation{Safran Tech, Digital Sciences and Technologies Department, Rue des Jeunes Bois, Châteaufort, 78114 Magny-Les-Hameaux, France}%
\author{Arnaud DERODE}
\affiliation{Institut Langevin, ESPCI Paris, CNRS, PSL University, 1 rue Jussieu, 75005 Paris, France}%
\author{Claire PRADA}
\affiliation{Institut Langevin, ESPCI Paris, CNRS, PSL University, 1 rue Jussieu, 75005 Paris, France}%

\date{\today}
\begin{abstract}
In a heterogeneous medium, the wavefield can be decomposed as an infinite series known as the Born expansion. Each term of the Born expansion corresponds to a scattering order, it is thus theoretically possible to discriminate single and multiple scattering contribution to the field. Experimentally, what is actually measured is the total field in which all scattering orders interfere. Conventional imaging methods usually rely on the assumption that the multiple scattering contribution can be disregarded. In a back-scattering configuration, this assumption is valid for small depths, and begins to fail for depths larger than the scattering mean-free path $\ell_s$. It is therefore a key issue to estimate the relative amount of single and multiple scattering in experimental data. To this end, a single scattering estimator $\hat{\rho}$ computed from the reflection matrix has been introduced in order to assess the weight of single scattering in the backscattered wavefield. In this article, the meaning of this estimator is investigated and a particular attention is given to recurrent scattering. In a diffraction-limited experiment, a multiple scattering sequence is said to be recurrent  if the first and last scattering events occur in the same resolution cell. Recurrent scattering is shown to be responsible for correlations between single scattering and higher scattering orders of the Born expansion, inducing a bias to the estimator $\hat{\rho}$ that should rather be termed confocal scattering ratio. Interestingly, a more robust estimator is built by projecting the reflection matrix in a focused basis. The argument is sustained by numerical simulations as well as ultrasonic data obtained  around 1.5~MHz  in a model medium made of nylon rods immersed in water. From a more general perspective, this work raises fundamental questions about the impact of recurrent scattering on wave imaging. 
\end{abstract}

\maketitle

\section{\label{sec:level1}Introduction}

As a wave propagates through a heterogeneous medium, it undergoes scattering: one part of its energy is diverted from the initial direction, and gives rise to secondary waves which in turn can be scattered again. Multiple scattering can be encountered with all kinds of waves, and has been a very active subject of research for several decades as well in quantum physics as in optics or acoustics~\cite{foldyMultipleScatteringWaves1945,ishimaruWavePropagationScattering1978,akkermansMesoscopicPhysicsElectrons2007,shengIntroductionWaveScattering2006,tiggelenWaveScatteringComplex2003,finkImagingComplexMedia2002,carminatiPrinciplesScatteringTransport2021}.

Imaging devices working in reflection as for radar echolocation or medical ultrasound, take advantage of single scattering in order to detect, locate, and possibly characterize the individual heterogeneities. Yet, in ultrasonic imaging, multiple scattering can be far from negligible, for instance in breast~\cite{aubryMultipleScatteringUltrasound2011} or liver~\cite{lambertReflectionMatrixApproach2020} tissues and can be even largely predominant in complex structures such as bones~\cite{Aubry2008} or lungs~\cite{Mohanty2017}. In the context of non-destructive evaluation, polycrystalline media like titanium alloys are intrinsically scattering media for ultrasonic waves due to the random orientations of crystallytes which generate a structural noise \cite{hirsekornScatteringUltrasonicWaves1982,yaldaPredictingUltrasonicGrain1996,wilcoxArrayImagingNoisy2011,kerbratImagingPresenceGrain2003,turnerElasticWavePropagation1999,weaverDiffusivityUltrasoundPolycrystals1990,vanpamelMethodologyEvaluatingDetection2014}. Defects can be detected by ultrasound, provided that the amount of multiple scattering between the grains is sufficiently low. If not, spikes on an ultrasound image might result in false alarms or, on the contrary, a defect might remain hidden in the clutter. Multiple scattering is thus a key issue since it causes conventional imaging techniques to fail. Therefore, whether it be for medical or non-destructive testing applications, there is a need for a depth-dependent indicator of the single scattering weight in the reflected wavefield.

In order to account for multiple scattering in a randomly disordered medium, the scattering mean-free path $\ell_s$ is a key parameter. In the case of an incoming plane wave propagating along the $z$-axis, the intensity of the ensemble-averaged wavefield $\left|\left\langle \psi\right\rangle  \right| ^2$ decays as $\exp(-z/\ell_s )$. Therefore the scattering mean-free path may be roughly thought of as a typical length scale to determine whether scattering has affected the incoming wave. If the path length is very large compared to $\ell_s$, the incoming wave loses its initial coherence while its energy is transferred to scattered waves; ultimately, the propagation of the average energy density can be described as that of classical particles undergoing a random walk, ruled by a diffusion equation.
 
Working at frequencies for which $\ell_s$ is large, for instance by lowering the frequency, is a classical way to diminish multiple scattering, usually at the cost of a poorer spatial resolution.
In the case of ultrasonic waves though, the advent of controllable multi-element arrays gave rise to alternative imaging methods involving a matrix approach. A matrix formalism is particularly appropriate since all the information available on the probed medium can be stored in the array response matrix or the so-called reflection matrix, which contains the set of impulse responses between each array element. Interestingly, single and multiple scattering were shown to exhibit different correlation properties in the reflection matrix measured on a random medium \cite{aubryRandomMatrixTheory2009,shahjahanRandomMatrixApproach2014}. Building on this difference, an algorithm was proposed to separate the single and multiple-scattering contributions to the reflection matrices. A first estimator of a multiple-to-single scattering ratio was built by Aubry \ea~\cite{aubryMultipleScatteringUltrasound2011} from the mean intensity of the reflection matrix diagonal elements ; then, Baelde \ea~\cite{baeldeEffectMicrostructuralElongation2018} proposed a single scattering estimator $\hat{\rho}$ using the Frobenius norms of the matrices. More recently, Lambert \ea~\cite{lambertReflectionMatrixApproach2020} and Velichko~\cite{velichkoQuantificationEffectMultiple2020} built other local multiple-to-single scattering ratio estimators calculated from the projection of the reflection matrix in a focused basis. In the aforementioned references, it is noticed that a residual multiple scattering term is not well separated from single scattering and thus remains in the single scattering estimated matrix. All these estimators are thus biased. The aim of this paper is to estimate and elucidate such bias by providing a physical analysis of the back-scattered echoes. In particular, we will show how part of the multiple scattering contributions, known as { \it recurrent scattering}, share common features with single scattering. In multiple scattering theory, a scattering sequence is said to be recurrent if the first and last scattering events occur at the same point. However, for receivers placed outside the medium, paths whose first and last scattering events take place in the same resolution cell also give rise to a long-range memory effect analogous to single scattering~\cite{aubryRecurrentScatteringMemory2014}, the former being just time delayed compared to the latter. In this paper, we will thus consider this last definition for recurrent scattering. 

Recurrent scattering has been the object of several studies, in particular, it was shown that recurrent scattering events do not contribute to the coherent backscattering enhancement~\cite{wiersmaExperimentalEvidenceRecurrent1995}; the role of recurrent scattering loops was also studied in the context of Anderson localization in a strong scattering regime~\cite{vollhardtDiagrammaticSelfconsistentTreatment1980,skipetrovDynamicsAndersonLocalization2006,aubryRecurrentScatteringMemory2014} ($k_0\ell_s \sim 1$, with $k_0$ the wavenumber). In this work, we investigate the impact of recurrent scattering on the reflection matrix properties in a much weaker scattering regime ($k_0\ell_s\gg1$). Recurrent scattering is shown to account for a bias made on the estimation of the single scattering component. The original single scattering estimator $\hat{\rho}$~\cite{baeldeEffectMicrostructuralElongation2018} can thus be re-interpreted as a confocal scattering ratio that quantifies the weight of single plus recurrent scattering in the reflection matrix.

Given the extreme variety and complexity of elastic wave propagation in biological or polycrystalline media \cite{huContributionDoubleScattering2015}, in this paper we choose to model much simpler media made of random distributions of isotropic scatterers with a numerical scheme based on the Born expansion. The advantage is that the total reflection matrix can be decomposed as a series of matrices  $\Knf$, $ n$  indicating the scattering orders,  that can be isolated and  separately investigated.    
The chosen numerical scheme also enables the discrimination of recurrent scattering paths among all possible multiple scattering paths. 

The paper is divided into four parts. The first section recalls the fundamentals of multiple scattering theory under Green's formalism. The second one explains the computation of the single scattering estimator $\hat{\rho}$; then, our analysis is applied to a proof-of-concept experiment that entails an assembly of parallel nylon wires embedded in water insonified by an ultrasonic linear array. The third section consists in translating the theory into a matrix formalism in order to predict the reflection matrix associated with a random distribution of scatterers, and compare the weight of both single and recurrent scattering to the experimental estimator $\hat{\rho}$. The dependence of the single plus recurrent scattering weight with respect to $\ell_s$ will be discussed. The last part shows the manifestation of recurrent scattering on the reflection matrix projected onto a focused basis in a generalized image space~\cite{lambertReflectionMatrixApproach2020,velichkoQuantificationEffectMultiple2020}. The impact of recurrent scattering on a local confocal scattering estimator is also discussed.

\section{\label{sec:level2}Born series and T-matrices}

In this section, we recall some basics of wave propagation in random media (more details can be found for instance in~\cite{frischWavePropagationRandom1968,bharucha-reidProbabilisticMethodsApplied2014,rytovPrinciplesStatisticalRadiophysics1989,ishimaruWavePropagationScattering1978}).  

In a homogeneous medium characterized by a wave velocity $c_0$, the scalar wave equation for the wave-field $\psi(\vec{r},t)$ writes 
\begin{equation}
\left(\Delta - \frac{1}{c_0^2} \frac{\partial^2 }{\partial t^2} \right) \psi(\vec{r},t) = s(\vec{r},t),
\end{equation}
with $s(\vec{r},t)$ the source distribution in the medium. In the case of a harmonic wave with angular frequency $\omega$, the associated Green's equation is 
\begin{equation}
\Delta G_0(\vec{r},\vec{r'},\omega) +k_0^2  G_0(\vec{r},\vec{r'},\omega) = \delta(\vec{r}-\vec{r'}),
\end{equation}
where $\delta$ is the Dirac distribution and $k_0=\omega/c_0$, the wave number. The homogeneous Green's function $G_0$ accounts for the propagation of a monochromatic wave between two points $\vec{r'}$ and $\vec{r}$ and reads :
\begin{equation}
 G_0 (\vec{r},\vec{r'},\omega) = \left\{ 
    \begin{array}{ll}
         -\frac{\imath}{4} \H0(k_0 \left| \vec{r} - \vec{r'} \right|) & \mbox{in 2D,}\\[6 pt]
         -\frac{\exp(\imath k_0 \left| \vec{r} - \vec{r'} \right|)}{4 \pi \left| \vec{r} - \vec{r'} \right|} & \mbox{in 3D,}
    \end{array}
\right.
\label{eq:chap2_solutionGreen}
\end{equation}
with $\H0$ the Hankel function of the first kind. In the far-field, the 2D Green's function can be approximated by:
\begin{equation}
	G_0(\vec{r},\vec{r'},\omega) \approx \frac{-e^{\imath \pi / 4}}{\sqrt{8 \pi k_0 \left| \vec{r} - \vec{r'} \right| }}  \exp \left( \imath k_0 \left| \vec{r} - \vec{r'} \right| \right).
	\label{eq:G0champlointain}
\end{equation}

In a heterogeneous medium where the wave speed varies randomly as a function of spatial coordinates, the Green's equation may be written as :
\begin{align} 
    \Delta G (\vec{r},\vec{r'},\omega) +k_0^2 \left(1-\mu(\vec{r})\right)G (\vec{r},\vec{r'},\omega)  = \delta(\vec{r}-\vec{r'}).
    \label{eq:GreenEqHetero}
\end{align} 
The non-dimensional quantity $\mu$ accounts for the medium heterogeneity. In the most common cases, it is a simple scalar: for instance, in optics, $1-\mu$ is the squared
refractive index $\left( c_0/c(\vec{r})\right) ^2$, $c_0$  being the speed of light in vacuum;  in acoustics,  $\psi$ being the pressure field, it is the same expression provided that the fluctuations of mass density at rest are ignored. With no loss of generality, Eq.~\eqref{eq:GreenEqHetero} also applies to less common cases: $\mu$ is then an operator and not a simple scalar~\cite{baydounRadiativeTransferAcoustic2016,baydounScatteringMeanFree2015}.

The solution of Eq.~\eqref{eq:GreenEqHetero} may be written recursively using the homogeneous space Green's function $G_0$:
\begin{equation}
\begin{split}
    G(\vec{r},\vec{r'},\omega)&= G_0(\vec{r},\vec{r'},\omega)  \\
    &+ k_0^2 \int  G_0(\vec{r},\vec{r_1},\omega) \mu(\vec{r_1}) 
G(\vec{r_1},\vec{r'},\omega) \dd\vec{r_1}.
\end{split}
\label{born_rec}
\end{equation}
Using the definition of the Green's function, the wavefield $\psi(\vec{r})$ is shown to follow the Lippman-Schwinger equation:
\begin{equation}
\psi(\vec{r})= \psi_0(\vec{r}) + k_0^2 \int  G_0(\vec{r},\vec{r_1},\omega) \mu(\vec{r_1}) \psi(\vec{r_1}) \dd\vec{r_1} ,
\label{lipp}
\end{equation}
with $\psi_0(\vec{r})$ the incident wavefield generated in the homogeneous medium by an arbitrary source distribution.

Both equations \eqref{born_rec} and \eqref{lipp} are recursive, and can be iterated to obtain the Born expansion, for the Green's function as well as for the wavefield: 
\begin{equation}
\begin{split}
   G(\vec{r},\vec{r'},\omega) &= G_0(\vec{r},\vec{r'},\omega) \\
   &+ k_0^2 \int  G_0(\vec{r},\vec{r_1},\omega) \mu(\vec{r_1})  G_0(\vec{r_1},\vec{r'},\omega) \dd\vec{r_1} \\
 &+ k_0^4 \iint  G_0(\vec{r},\vec{r_2},\omega) \mu(\vec{r_2})  G_0(\vec{r_2},\vec{r_1},\omega) \mu(\vec{r_1}) \\ & \qquad \qquad \qquad \qquad \qquad G_0(\vec{r_1},\vec{r'},\omega) \dd\vec{r_1} \dd\vec{r_2} \\ &+ \dots 
\label{born}
\end{split}
\end{equation}
and,
\begin{equation}
\begin{split}
\psi(\vec{r}) &= \psi_0(\vec{r}) \\
&+ k_0^2 \int  G_0(\vec{r},\vec{r_1},\omega) \mu(\vec{r_1}) \psi_0(\vec{r_1}) \dd \vec{r_1} \\
&+ k_0^4 \iint  G_0(\vec{r},\vec{r_2},\omega) \mu(\vec{r_2})  G_0(\vec{r_2},\vec{r_1},\omega) \mu(\vec{r_1}) \\ & \qquad \qquad \qquad \qquad \psi_0(\vec{r_1}) \dd \vec{r_2} \dd\vec{r_1} \\ &+ \dots
\label{bornPsi}
\end{split}
\end{equation}
Neither \eqref{born_rec} nor \eqref{lipp} are explicit solutions of the scattering problem, they are just a recursive expression of the solution. The Born development is an actual expression of the solution, but it contains an infinite number of terms, corresponding to various scattering orders.

Let $V$ be such that $V\left(\vec{r}_2,\vec{r}_1 \right) =k_0^2\mu(\vec{r}_1)\delta(\vec{r}_2-\vec{r}_1)$. Then for the sake of brevity, matrix products can be used instead of multiple integrals, for instance:
\begin{equation}
  \mathbf{V G_0 V} \leftrightarrow \iint V(\vec{r_2},\vec{x}) G_0(\vec{x},\vec{y})V(\vec{y},\vec{r_1}) \dd\vec{x} \dd\vec{y}.
\end{equation}
Defining the $\mathbf{T}$-matrix (or scattering matrix) as $\mathbf{T}= \mathbf{V}+\mathbf{VG_0V}+\mathbf{VG_0VG_0V}+\dots$ yields an explicit expression for the total wavefield (incident + scattered):
\begin{equation}\label{eq:defT}
\psi(\vec{r})= \psi_0(\vec{r}) + \iint G_0 (\vec{r},\vec{r_2},\omega) T(\vec{r_2},\vec{r_1},\omega) \psi_0(\vec{r_1}) \dd\vec{r_2} \dd \vec{r_1}.
\end{equation}
In Eq.~\eqref{eq:defT}, the entire complexity of the medium is wrapped up in $\mathbf{T}$. Eq.~\eqref{eq:defT} can be interpreted as the following series of events: the incident wave $\psi_0$ impinges at $\vec{r_1}$, is affected by $\mathbf{T}$, exits at $\vec{r_2}$, then the resulting wave propagates freely to the receiver at $\vec{r}$. In Eq.~\eqref{eq:defT}, $\mathbf{T}$  is the scattering matrix of the entire medium, as if it was one single large scatterer: intrinsically there is no ``multiple scattering between two points'', unlike in the Born development (Eq.~\eqref{bornPsi}). \\

For a medium composed of $N_s$ discrete objects embedded in a homogeneous fluid, it is common to adopt an intermediate scale, and consider each object, even if it is not point-like, as the unit scattering cell. Let $\mathbf{T}_i$ denote the $\mathbf{T}$-matrix of the $i^{th}$ object; then, Eq.~\eqref{eq:defT} can be developed into a series of scattering sequences:
\begin{equation}
\begin{split}
\mathbf{\psi} &=  \mathbf{\psi}_0 + \sum_{i=1}^{N_s} \mathbf{G}_0 \mathbf{T}_i \mathbf{\psi}_0  + \sum_{i=1}^{N_s} \sum_{\substack{j=1 \\ j\ne i}}^{N_s} \mathbf{G}_0 \mathbf{T}_i \mathbf{G}_0 \mathbf{T}_j \mathbf{\psi}_0 + \dots
\end{split}
\label{tmatrix}
\end{equation}
The terms on the right-hand side of Eq.~\eqref{tmatrix} correspond respectively to the incident field, the single-scattering contribution, the double-scattering contribution, etc. The total wavefield $\psi$ may be  decomposed in that manner whatever the chosen unit scattering cell.  In the following, Eq.~\eqref{tmatrix} is the fundamental relation that will be used to compute recursively the reflection matrix, considering one cylindrical scatterer as the unit scattering cell.

\section{\label{sec:level2b} The single scattering ratio estimation}

\begin{figure}[hbt!]
	\centering
	\includegraphics[width = 8.6cm]{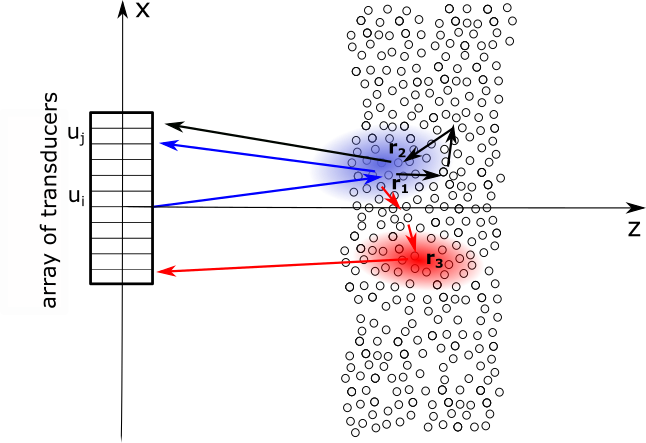}
	\caption{
	Experimental configuration: an ultrasonic probe is used to record the reflection matrix $\mathbf{K}=[K(u_i,u_j,t)]$ associated with the scattering medium. The incident wave emitted by one transducer at $\vec{u_i}$ undergoes a first scattering event at $\vec{r_1}$. The wave can then either go back directly towards the transducer $\vec{u_j}$ (single scattering, blue path) or be multiply-scattered. Multiple scattering paths can be classified in two categories: (\textit{i}) recurrent scattering paths whose first and last scattering events at $\vec{r_1}$ and $\vec{r_2}$ occur within overlapping resolution cells (black arrows); (\textit{ii}) non-recurrent scattering paths (red arrows).}
	\label{fig:milieu} 
\end{figure}

Let us consider the experimental setup in Fig.~\ref{fig:milieu}. An array of $N$ emitter-receivers at positions $\vec{u_i}=(u_i,0)$, ($i=1,\dots,N$) is placed in front of the scattering medium under investigation. The $N\times N$ inter-element impulse responses between all possible transducers  are measured. A Fourier transform yields the reflection matrix \mbox{$\Ktf=[K(u_i,u_j,\omega)]$} at each angular frequency $\omega$. In actual experiments, the various scattering orders described by Eq.~\eqref{tmatrix} cannot be discriminated in $\Ktf$.
Nevertheless, one can try to isolate the single scattering contribution from the other terms. To that aim, Aubry \ea~\cite{aubryRandomMatrixTheory2009,Aubry2009} proposed to apply a matrix manipulation based on the peculiar correlation of the single scattering matrix $\mathbf{K}^{(1)}$. We first recall the principle of this method, which will be referred to as the matrix filter in the following, we then define the single scattering estimator \cite{baeldeEffectMicrostructuralElongation2018} and apply it on a model experiment. 

\subsection{\label{sec:level2b_1}Theoretical derivation}
\label{sec:filter}

 The  expression of the single scattering matrix $\Kf$ is the second term of Eq.~\eqref{tmatrix}. For scatterers smaller than the wavelength, the $\mathbf{T}$-matrix of the s$^{th}$ scatterer located at position $r_s=(x_s,z_s)$ can be written \mbox{$T_s(\vec{r},\vec{r^\prime},\omega) = \Ts(\omega) \delta(\vec{r}-\vec{r_s}) \delta(\vec{r^\prime}-\vec{r_s})$}, with $\Ts(\omega)$, the scatterer frequency response. Moreover, when considering the emitter and receiver sizes also much smaller than the wavelength, $\mathbf{\psi}_0$ can be replaced by the Green's matrix $\mathbf{G}_0$ that contains the $N \times N_s$ free space Green's functions between the transducers and the scatterers. Thus, the coefficients of $\Kf$ can be expressed as follows 
\begin{equation}
   	K^{(1)}(u_i,u_j,\omega)
    =	\sum\limits_{s=1}^{N_s}G_0(\ui,\vec{r_s},\omega) \Ts(\omega)  G_0(\vec{r_s},\uj,\omega).
	\label{eq:K1int}
\end{equation}

In a time-resolved experiment, the received waves within a given time-window $[T-\Delta T/2;T+\Delta T/2]$ come from a specific region, called the isochronous volume $\Gamma(T,\Delta T)$~\cite{mallartAdaptiveFocusingScattering1994}. In the single scattering regime, $\Gamma(T,\Delta T)$ is the locus of points $\vec{r}$ such that $\left| \vec{u_i}-\vec{r}\right|+\left| \vec{r}- \vec{u_j}\right|\in[2z-2\Delta z;2z+2\Delta z]$, with $z=c_0T/2$ and $\Delta z=c_0 \Delta T/2$, which describes a skein of ellipses. In the far-field, $\Gamma(T,\Delta T)$ can be approximated by a slice parallel to the transducers array, between depths $z\pm  \Delta z/2$. It is therefore a common procedure in ultrasound imaging to time-gate the reflected signals, then analyze their properties as a function of the central time $T$ (or the equivalent depth $z=c_0T/2$). It should be noted  that as long as $2z$ is larger than the transverse size of the array, the direct path between the emitter and the receiver is eliminated by time-gating. And in the expression $\Kf$, only the scatterers contained in the isochronous volume $\Gamma$ are considered.

Injecting Eq.~\eqref{eq:G0champlointain} into Eq.~\eqref{eq:K1int} and considering the isochronous volume $\Gamma$ as a thin slice of scattering medium around depth $z$, the elements of $\Kf$ can be written as follows:
\begin{equation}
	\begin{split}
K^{(1)}(u_i,u_j,\omega) \propto \sum\limits_{(x_s,z_s)\in\Gamma} \Ts(\omega) &\exp \left(\imath k_0 \sqrt{z_s^2+(x_s-u_i)^2} \right) \\ &\exp \left(\imath k_0 \sqrt{z_s^2+(x_s-u_j)^2} \right).
	\end{split}
\end{equation}
Under the hypothesis of the paraxial approximation \mbox{$k_0(x_s-u_i)^4/(8z_s^3)<<\pi$}, this expression can be rewritten as follows
\begin{equation}
	\begin{split}
			K^{(1)}(u_i,u_j,\omega) \propto  & \exp \left( \imath k_0 \frac{(u_i-u_j)^2}{4z} \right) \\ \sum\limits_{(x_s,z_s)\in\Gamma}e^{\imath k_0(z_s-z)}\Ts(\omega)&\exp \left(\imath k_0 \frac{(u_i+u_j-2x_s)^2}{4z} \right).
	\end{split}
	\label{parax}
	\end{equation}

The first phase term is deterministic and is the same for all emitter-receiver pairs $(i,j)$ such that $u_i-u_j$ is constant i.e., along the diagonals of $\Kf$. The second term depends on the scatterers configuration, therefore it is random and changes from one realization to the next. However it is a function of {$(u_i+u_j)$} hence it is constant along a given anti-diagonal of $\Kf$. As a consequence, the single scattering matrix $\Kf$ exhibits a deterministic coherence along its anti-diagonals, also known as the memory effect, which can be expressed in the following manner: 

\begin{equation}
    K^{(1)}(u_{i-d},u_{i+d},\omega)=K^{(1)}(u_i,u_i,\omega)\exp \left(\imath k_0 \frac{(u_{i-d} - u_{i+d})^2}{4z} \right).
    \label{eq:coherence}
\end{equation}

There are $N \times N$ matrix elements $K^{(1)}(i,j)$, hence $2N-1$ anti-diagonals than can be labelled from \mbox{$l=1$} to \mbox{$l=2N-1$}. Following~\cite{baeldeEffectMicrostructuralElongation2018}, we define:
\begin{equation}
	E_l(i,j, \omega,T)= \left\{
	\begin{array}{ll}
		0 \text{ if } i+j \neq l+1, \\
		\\
		\dfrac{\exp \left(\imath k_0 \dfrac{(u_i-u_j)^2)}{4z} \right)}{\sqrt{\text{min}(l,2N-l)}} \text{ if } i+j = l+1.
	\end{array}
	\right.
	\label{ds}
\end{equation}
The next operation consists in projecting the reflection matrix $\Ktf$ onto the ``single scattering space'' generated by the set of matrices $\left\{ \En \right\}_{1 \leq l \leq 2N-1}$. At each depth and each frequency, we obtain a $N \times N$ ``filtered'' matrix denoted $\Kfilt$: 
\begin{equation}
	\Kfilt = \sum_{l=1}^{2N-1} \left\langle \En | \Ktf \right\rangle \En.
	\label{eq:proj}
\end{equation}
By doing so, we select the part of $\Ktf$ that follows coherence along anti-diagonals as stated in Eq.~\eqref{eq:coherence}. 
Next, we introduce the following estimator:
\begin{equation}
	\hat{\rho} = \frac{\norme{\Kfilt}^2  }{\norme{\Ktf}^2 }.
	\label{eq:rhof}
\end{equation} 
 This estimator can be studied as a function of time (or equivalent depth $z=c_0T/2$) as well as angular frequency $\omega$.  It can  be   averaged over the frequency band, or  over realizations of disorder, which can be achieved by randomly drawing a large number of configurations.

If the matrix filter worked ideally, we should have \mbox{$\Kfilt = \Kf$}, hence $\hat{\rho}$ could be interpreted as the proportion of single scattering within the reflected wave field; that would be a valuable quantity which could be studied as as function of depth $z$ and angular frequency $\omega$.  Before going deeper into the theory, we provide an experimental illustration of this estimator. 

\subsection{\label{sec:level2b_2}Experimental results}

The single scattering estimator is now applied to ultrasonic measurements carried out in a water tank. The scattering medium consists of a collection of randomly distributed parallel nylon wires of radius $a$= 0.1 mm. The wire sample is of dimensions 150$\times$135 mm, with a concentration of 4 wires.cm$^{-2}$, so that  the fractional density is approximately 0,125\%. Longitudinal and transverse wave speeds in nylon are 2500 m.s$^{-1}$ and 1100 m.s$^{-1}$ respectively. The density of nylon being close to the one of water, the density contrast can be neglected and the scattering considered as isotropic at low frequencies. This was confirmed in a previous study by Minonzio \ea~\cite{minonzioDecompositionOperateurRetournement2006a} who calculated  and measured the reflection matrix of a single wire and derived its decomposition   onto the cylindrical normal modes of vibration. 
Measurements were done with a linear array (Imasonic, Besançon, France) composed of $N=$ 64 piezoelectric transducers with pitch $p=$ 0.5 mm and 1.5 MHz central frequency. The distance $z_0$ between the probe and the front face of the sample is 140 mm. To acquire the reflection matrix, each element is excited by a  chirp of 30 µs duration  and  0.5-2.5 MHz bandwidth  with a von Hann window apodization. The experimental setup is sketched in Fig.~\ref{fig:milieu}.
\begin{figure}[hbt!]
\centering

\includegraphics[width=8.6cm]{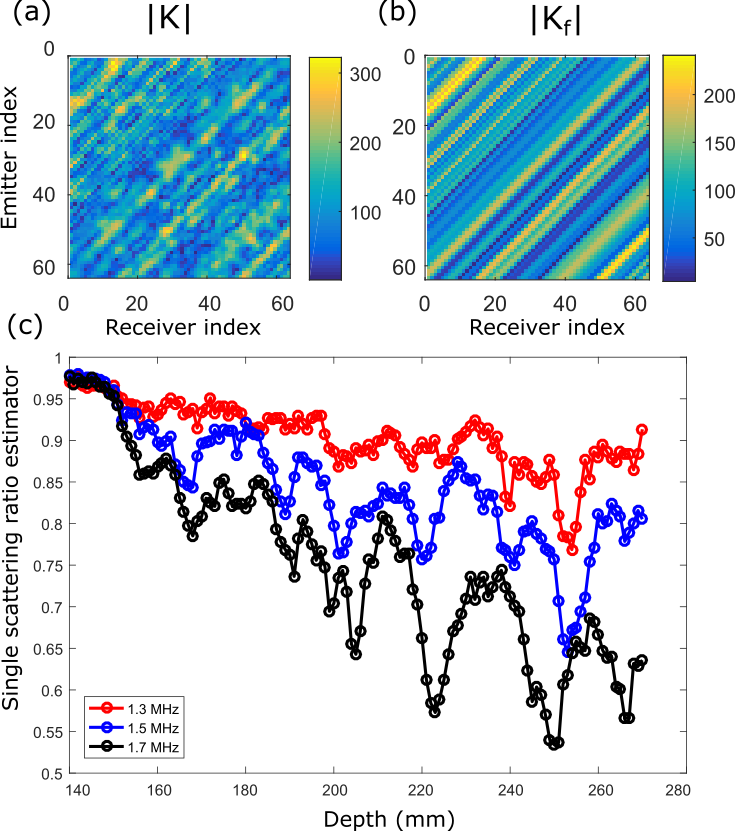}
\caption{\label{fig:propDS_exp} Results obtained from measured data on a nylon wire sample for 
 and 
$\Delta T =$ 10 µs: (a) reflection matrix at 1.5 MHz for depth $z=205$ mm, (b) filtered matrix obtained from (a), (c) experimental single scattering ratio estimator at 1.3 MHz, 1.5 MHz and 1.7 MHz obtained for one realization.}
\end{figure}
 
As a typical example, Fig.~\ref{fig:propDS_exp} displays the reflection matrices before ($\Ktf$) and after ($\Kfilt$) applying the matrix filtering process described in Sec.~\ref{sec:filter}. $\Ktf$ and $\Kfilt$ are very different, which indicates the existence of multiple scattering. The estimator of the single scattering ratio is plotted on Fig.~\ref{fig:propDS_exp}(c), as a function of depth $z$, at 1.3, 1.5 and 1.7 MHz. As expected, the slopes of the curves increase with frequency, which means that the medium exhibits stronger multiple scattering with increasing frequency (smaller scattering mean-free path).

One purpose of this article is to analyze and use the information contained in the filtered matrix, and in particular, to understand how  $\Kfilt$ differs from the theoretical single scattering matrix $\Kf$. In the next section, we present a numerical calculation of the successive scattering terms $\Knf$ ($n=1, \dots,\infty$) contributing to the total reflection matrix $\Ktf$. Then, the true single scattering ratio \mbox{$\rho=\norme{\Kf}^2/\norme{\Ktf}^2$} can be computed and compared to its estimator $\hat{\rho}$.

\section{\label{sec:level3}Born expansion of the reflection matrix}

In this section, we describe the theoretical computation of the reflection matrix for a random distribution of scatterers using Born expansion. A similar approach was used by Minonzio \ea~\cite{minonzioDecompositionOperateurRetournement2006a} and then by Fan \ea~\cite{fanComparisonUltrasonicArray2014} to investigate the effect of multiple scattering of two scatterers in a homogeneous medium on imaging algorithms. The analytical expression of double scattering by a random distribution of scatterers was then analytically derived by Hu and Turner \cite{huContributionDoubleScattering2015}. In this article, for each realisation of disorder, all scattering orders are taken into account. \\

\subsection{\label{sec:level3_1}The scattering medium}

The numerical scheme is two-dimensional. The scattering medium is a cloud of $N_s$ identical non-overlapping cylinders of radius $a$, embedded in a fluid of sound velocity $c_0$ and randomly distributed on a rectangular area of surface $A$. It is assumed that the density contrast between the cylinders and surrounding medium is negligible, so that heterogeneity comes from the compressibility contrast; only longitudinal (pressure) waves are taken into account. Denoting $c_s$ the velocity inside the cylinder, the random term $\mu$ in the wave equation \eqref{eq:GreenEqHetero} is \mbox{$\mu=1-(c_s/c_0)^2$} inside the scatterers and \mbox{$\mu=0$} outside. The radius $a$ is smaller than the wavelength, so the differential scattering cross section of a cylinder is isotropic. Under these conditions, the frequency response $\T(\omega)$ of each scatterer derived in Appendix \ref{appendix:tmatrix}  is given by
\begin{equation}
	\T(\omega) = k_0^2 \frac{\pi a^2 \mu}{1 + \frac{\imath k_0^2}{4}\pi a^2 \mu}. 
\end{equation}
and the scattering cross-section of a single scatterer is given by~
\begin{equation}
	\sigma (\omega) = \frac{k_0^3}{4}\left| \frac{\pi a^2 \mu}{1 + \frac{\imath k_0^2}{4}\pi a^2 \mu} \right|^2 .
	\label{eq:sigma}
\end{equation} 
Under the independent scattering approximation, the scattering mean-free path is:
\begin{equation}
	\ell_s (\omega) = \frac{1}{n_s \sigma (\omega)},
	\label{eq:ls}
\end{equation}
with $n_s=N_s/A$ the number of scatterers per unit surface.

\subsection{\label{sec:level3_2}Reflection matrices}

At each angular frequency $\omega$, the $N\times N$ reflection matrix  \mbox{$\Ktf=[K(u_i,u_j,\omega)]$} may be decomposed as a Born expansion, with all scattering orders from $n=1$ to infinity:
\begin{equation}
	\Ktf=\sum_{n=1}^\infty \Knf  ,
	\label{eq:somme}
\end{equation}
with $\Knf$, the n$^{th}$ scattering order of the reflection matrix.

The single scattering matrix $\Kf$ can be expressed by rewriting Eq.~\eqref{eq:K1int} under a matrix formalism:
\begin{equation}
	\Kf(\omega) = \mathbf{G}_0(\omega) \times \T(\omega) \times \mathbf{G}_0^\top(\omega).
	\label{eq:K1}
\end{equation}
$\mathbf{G}_0$ denotes the $N \times N_s$ matrix whose elements are the Green's function $G_0(\ui,\vec{r_s})$ between each array element and each scatterer. To avoid dimensional confusion,  $\G$ denotes the $N_s \times N_s$ matrix whose elements are the Green's function $G_0(\vec{r_s},\vec{r_p})$ between two scatterers. Accordingly, higher scattering orders $\Knf$ ($n>1$) can be deduced from Eq.~\eqref{tmatrix}:
\begin{equation}
	\Knf(\omega) =\mathbf{G}_0(\omega)\times \T(\omega) \times (\G(\omega)\T(\omega))^{n-1} \times \mathbf{G}_0^\top(\omega).
	\label{eq:Kn}
\end{equation}
Injecting  Eq.~\eqref{eq:K1} and Eq.~\eqref{eq:Kn} into Eq.~\eqref{eq:somme} leads to the following expression of the reflection matrix:
\begin{equation}
	\Ktf(\omega) =  \mathbf{G}_0(\omega)\times \T(\omega) \times \left( \Id-\T(\omega)\G(\omega) \right)^{-1} \times \mathbf{G}_0^\top(\omega),
	\label{eq:Ktot}
\end{equation}
where the exponent $^{-1}$ denotes matrix inversion and $\Id$ the identity matrix. \\

The condition of convergence of the geometrical series $ \sum \Knf$ is ${\left| \nu\right|  < 1}$, with $\nu$ any eigenvalue of  $\G \T$~\cite{osnabruggeConvergentBornSeries2016}. Note that if ${\left| \nu \right| \geqslant 1}$, only the expression of $\Ktf$ given by Eq.~\eqref{eq:Ktot} has a 
physical meaning. In this paper, as we are interested in the different scattering orders, we will only consider converging cases.\\

\subsection{\label{sec:level3_3}Numerical results}

Simulation parameters are chosen to be as close as possible to the experimental configuration described in Sec.~\ref{sec:level2b}. We considered an array of $N=$ 64 transducers with central frequency \mbox{1.5 MHz}, 1 MHz bandwidth with a von Hann apodisation and pitch \mbox{$p$ = 0.5 mm}. \mbox{$N_s$ = 810 scatterers} with velocity \mbox{$c_s$ = 2500 m.s$^{-1}$} are placed in a rectangular area delimited by depths between 140 mm and 275 mm and off-axis distances below 75 mm. The density is 4 scatterers/cm$^2$; the radius is $a=0.1$ mm. The ambient fluid is water (\mbox{$c_0$ = 1480 m.s$^{-1}$}). The scattering mean-free path $\ell_s$ derived from Eq.~\eqref{eq:ls} is 970 mm at 1.5 MHz which is much larger than the sample thickness $L=135$ mm. This medium is thus weakly scattering.\\

\begin{figure}[hbt!]
	\centering
	\includegraphics[width =8.6cm]{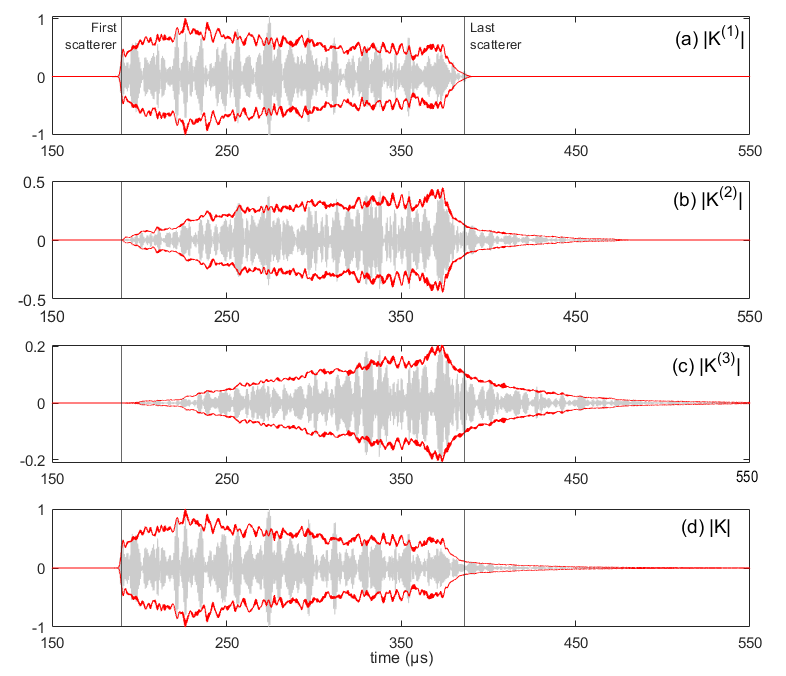}
	\caption{Numerical results: impulse responses computed from (a) the single scattering matrix $\Kf$, (b) the double scattering matrix $\Kdf$, (c) the triple scattering matrix $\Ktrf$, (d) the total scattering matrix $\Ktf$. The red curves are the mean envelope of the impulse responses, averaged over all emitter-receiver pairs. }
	\label{fig:ascan}
\end{figure}

A 1-D inverse Fourier Transform of $\Ktf(\omega)$, $\Kf(\omega)$ and $\Knf(\omega)$ defined by equations \eqref{eq:K1} to \eqref{eq:Ktot} provide the reflected signals in the time domain. Examples of impulse responses $K^{(n)}(u_i,u_j,t)$ corresponding to different scattering orders are shown in Fig.~\ref{fig:ascan}. As expected, the multiple scattering contributions slowly increase after the first arrivals and persist after the end of single scattering signal, resulting in a coda whose duration increases with the scattering order. The shortest distance between a scatterer and the array elements is 140 mm, and the largest is 294 mm; as a result the single scattering contributions occurs between 190 µs and 390 µs ($t=0$ is the emission time). The  scattering mean-free time is $\ell_s/c_0=650$ µs at 1.5 MHz; this order of magnitude is consistent with the fact that the double scattering contribution increases with time and its amplitude becomes comparable to that of single scattering a few tenths of microseconds after the arrival of the front face echo.
A short-time Fourier analysis is applied to impulse responses, which yields one complex-valued $N \times N$  matrix at each time $T$ and frequency $\omega$. 10-µs time-windows (6 to 7 periods) are used, so that if the single scattering and far-field approximations were valid, the isochronous area $\Gamma$ would correspond to a 7.5 mm-thick slice of the scattering medium. To avoid heavy notations, the time or frequency dependence of matrices will be omitted in the following.
  
\begin{figure*}[hbt!]
	\centering
	\includegraphics[width=17.2cm]{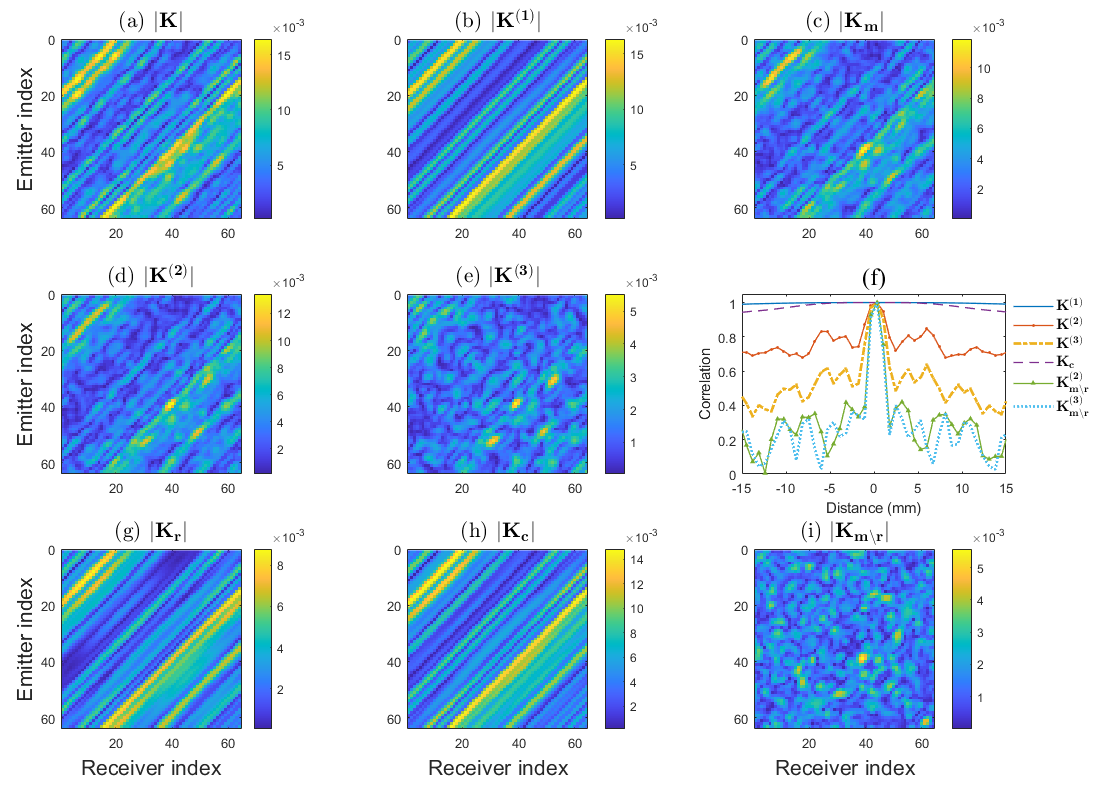}
	\caption{Modulus of the elements of the different calculated matrices : (a) total scattering matrix $\Ktf$, (b) single scattering matrix $\Kf$, (c) multiple scattering matrix $\Kmult = \Ktf - \Kf$, (d) double scattering matrix $\Kdf$, (e) triple scattering matrix $\Ktrf$, (g) recurrent scattering matrix $\Krec$, (h) confocal scattering matrix $\Kc = \Kf + \Krec$, (i) conventional multiple scattering matrix $\Kpmult = \Kmult - \Krec$. Matrices are time-gated (10 µs window around depth 205 mm) and shown at the central frequency. (f) correlation along the anti-diagonals as a function of the distance to the main matrix diagonal (Eq.~\eqref{eq:correlation}).}
	\label{fig:matrices}
\end{figure*}

Examples of time-gated matrices for different scattering orders are presented in Fig.~\ref{fig:matrices}. As expected, the single scattering matrix $\Kf$ displays a long-range coherence along its antidiagonals Fig.~\ref{fig:matrices}(b). More surprisingly, this coherence seems to persist, at least partially, for the second and third scattering order matrices,  $\Kdf$ (Fig.~\ref{fig:matrices}(d)) and $\Ktrf$ (Fig.~\ref{fig:matrices}(e)). It results in a multiple scattering matrix $\Kmult=\Ktf - \Kf$ (Fig.~\ref{fig:matrices}(c)) that is far from having uncorrelated elements contrary to the assertions given in previous works \cite{aubryMultipleScatteringUltrasound2011,baeldeEffectMicrostructuralElongation2018}. Consider matrix elements $(u,u)$ and $(u-\Delta u, u+\Delta u)$: the former is on the main diagonal, the latter is on the same anti-diagonal. The persistence of memory effect (cf. property Eq.~\eqref{eq:coherence}) in the n$^{th}$ multiple scattering matrix can be measured by the correlation coefficient:
\begin{equation}
      C^{(n)}(\du) \nonumber = \langle K^{(n)}(u,u)K^{(n)*}(u-\du,u+\du)\rangle_u ,
    \label{eq:correlation}
\end{equation} 
and its normalized version,
\begin{eqnarray}
      & &  \hat{C}^{(n)}(\du) \nonumber \\ 
         &=& \frac{ C^{(n)}(\du) }{\sqrt{\langle |K^{(n)}(u,u)|^2\rangle_u \langle |K^{(n)}(u-\du,u+\du)|^2\rangle_u}},
    \label{eq:correlation_norm}
\end{eqnarray} 
where the symbol $\langle \cdots \rangle_u$ denotes an average over each diagonal. $\hat{C}^{(n)}(\du)$ evaluates the degree of correlation between anti-diagonal elements $(u-\Delta u, u+\Delta u)$ of matrices $\Knf$. \\

Figure~\ref{fig:matrices}(f) shows the correlation coefficient associated with matrices $\Kf$, $\Kdf$ and $\Ktrf$. A constant correlation coefficient $\hat{C}^{(1)}(\du)$ is found for the single scattering component: this is the manifestation of the long-range memory effect highlighted by Eq.~\eqref{parax}. Interestingly, the correlation coefficients $\hat{C}^{(2)}(\du)$ and $\hat{C}^{(3)}(\du)$ display the following shape: a narrow peak characteristic of 
{a multiply-scattered (\textit{i.e} spatially-incoherent) wave-field and of shape equal to the coherent back-scattering peak~\cite{Tourin1997,aubryCoherentBackscatteringFarfield2007} (see Appendix~\ref{appC})}, 
on top of a constant background characteristic of a long-range correlation, similar to the memory effect exhibited by $\Kf$. 
In other words, the correlation described by Eq.~\eqref{eq:coherence} that was typical of single scattering also appears in the  double-scattering contribution $\Kdf$, and also (yet to a lesser level) in the triple-scattering contribution $\Ktrf$. In the light of this surprising result, the validity of $\hat{\rho}$ as an indicator of single scattering ratio is investigated in the next paragraph.

\begin{figure}[hbt!]
	\centering
\includegraphics[width =8.6cm]{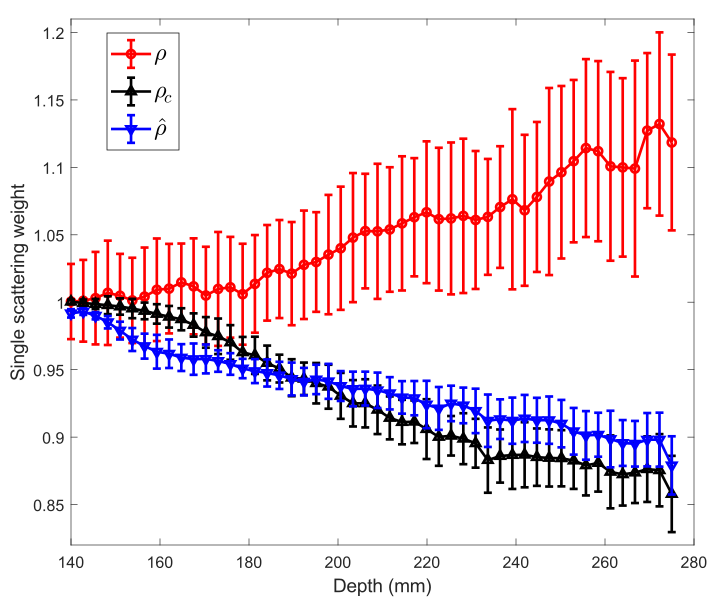}
	\caption{\label{fig:propDS} Average and standard deviation over 50 medium realizations of the single scattering ratio $\rho$, the confocal scattering ratio $\rho_{c}$ and the estimator $\hat{\rho}$ with a 10 µs time-gating.}
\end{figure}

\begin{figure}[hbt!]
	\centering
	\includegraphics[width =8.6cm]{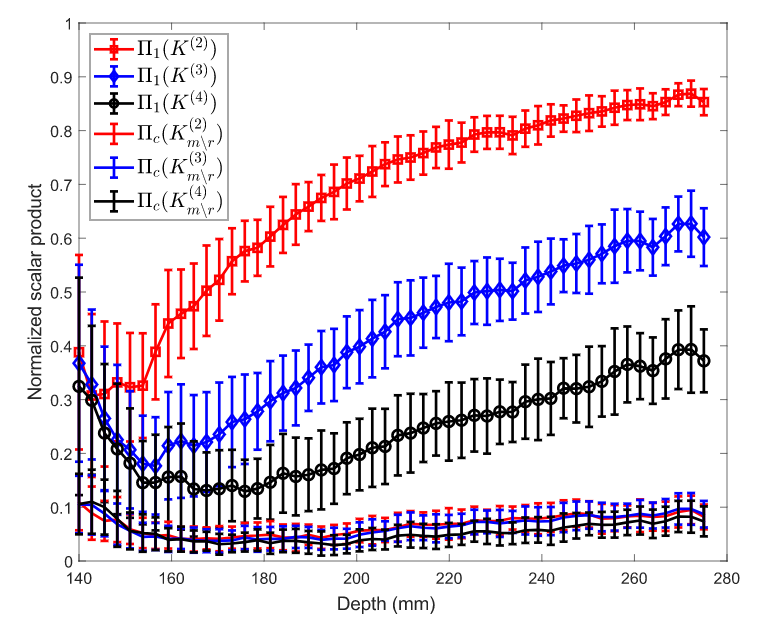}
	
\caption{\label{fig:ps} Average and standard deviation over 50 medium realizations of the normalized scalar products $\Pi_1(\Knf)$ (Eq.~\eqref{eq:pi}) and $\Pi_c(\Kpn)$ (Eq.~\eqref{eq:pir}).}
\end{figure}

\subsection{\label{sec:level3_4}Single scattering ratio estimator and first order Born approximation.}

 The single scattering ratio estimator $\hat{\rho}$ averaged over 50 realizations of disorder is plotted  in Fig.~\ref{fig:propDS} as a function of time (blue line). As observed experimentally in Fig.~\ref{fig:propDS_exp}, $\hat{\rho}$ decreases with time. The orders of magnitude for the decay length of $\hat{\rho}$ are in reasonable agreement with the experimental results:  at depth 260 mm, $\hat{\rho} \approx$ 0.85 in the experiment while $\hat{\rho} \approx 0.9$ in the simulation.  The discrepancies  between  simulated and  experimental curves computed at the same frequency (1.5 MHz, blue curve on Fig.~\ref{fig:propDS_exp}) can be explained by the simplified model used in the  simulation. First, the diffraction in the third dimension that occurs in the experiment is not taken into account by the model; second, the simulated $\mathbf{T}$-matrix  does not take into account shear velocity and density contrast and considers monopolar scattering only. As a consequence, the numerical scattering cross-section of the scatterers does not precisely agree with the experimental one. 

The numerical scheme yields a straightforward definition of the true single scattering ratio as the ratio between the norms of the single and total scattering matrices:
\begin{equation}
\rho=  \frac{\norme{\Kf}^2}{\norme{\Ktf}^2}.
\label{eq:rho1}
\end{equation}

The single scattering ratio  ${\rho}$ averaged over 50 realizations of disorder is plotted  in Fig.~\ref{fig:propDS} as a function of time (red circles). Surprisingly, unlike the estimator $\hat{\rho}$, the single scattering ratio $\rho$ is found to be larger than 1 and even increasing with time.  The fact that $\norme{\Kf} > \norme{\Ktf}$ may seem counter-intuitive; going back to the definition of $\Ktf$ as the sum of the scattering orders, this is possible only if  $\Kf$ and  higher scattering order matrices are correlated in such a way that the norm of their sum is not equal to the sum of their individual norms. To investigate this, we proposed to compute the normalized scalar product $\Pi_1\left (\Knf \right ) $ between the single and the $n^{th}$ order scattering matrices, $\Kf$ and $\Knf$. At each frequency and time, $\Pi_1\left (\Knf \right )$ is computed as follows: 
\begin{equation}
\Pi_1\left (\Knf \right ) = \frac{|\left\langle \Knf | \Kf \right\rangle|}{\norme{\Knf}\norme{\Kf}}.
\label{eq:pi}
\end{equation}
Fig.~\ref{fig:ps} displays $\Pi_1(\Knf) $ as a function of depth for $2 \leq n \leq 4$. We observe that the correlation between $\Knf$ and $\Kf$ increases with depth and decreases with the scattering order, however this correlation remains strong up to the fourth scattering order. This is consistent with the observation made on the multiple scattering matrices (Fig.~\ref{fig:matrices}(c) to Fig.~\ref{fig:matrices}(e)) and the correlation coefficient of Eq.~\eqref{eq:correlation} (Fig.~\ref{fig:matrices}(f)). It explains why $\Kfilt$ cannot always be a perfect estimator of $\Kf$: the ``matrix filter'' does not extract single scattering in the sense of the first order Born approximation. Indeed it extracts all contributions (including some of multiple scattering) which entail the memory effect of Eq.~\eqref{eq:coherence}. In the next section, we will show how to build new matrices with less correlation between the scattering orders and emphasize the weight of recurrent scattering on $\hat{\rho}$.

\section{\label{sec:level4}Recurrent scattering}

As explained in Sec.~\ref{sec:level2b}, the single scattering matrix  $\Kf$ (Fig~\ref{fig:matrices}(b)) exhibits a spatial coherence along the anti-diagonals, known as the memory effect in
optics~\cite{freundMemoryEffectsPropagation1988}, which can be taken advantage of to discriminate single from multiple scattering contributions~\cite{shahjahanRandomMatrixApproach2014}. Surprisingly, the multiple scattering contribution $\Kmult$ exhibits the same kind of anti-diagonal correlation, though to a lesser degree. In this section, we provide a theoretical analysis of this phenomenon which we explain as a consequence of the combined effects of recurrent scattering and diffraction-limited resolution. Then, to confirm this analysis, we use the numerical scheme to discriminate recurrent and non-recurrent multiple scattering contributions.

\subsection{\label{sec:level4_1} Antidiagonal coherence in $\Kmult$}

A multiple scattering sequence involves at least two distinct scatterers. Let us denote $\ri$ and $\rj$ their positions. Taking the individual scatterer as the unit scattering cell, we sum all over possible entry and exit scatterer pairs whose positions are compatible with the time-gating condition. Using the properties of the $\mathbf{T}$-matrices, the $(i,j)$ element of $\Kmult$ may be written as: 
\begin{equation}
\begin{split}
\label{Km}
      K_m(u_i,u_j) =   k_0^2 \iint\limits G_0(\ui,\ri) & \mu(\ri) G(\ri,\rj) \mu(\rj)\\ &G_0(\rj,\uj) \dd\ri\dd\rj.
\end{split}
\end{equation}
Only the multiple scattering sequences whose path lengths are comprised between $c_0(T-\Delta T/2)$ and $c_0(T+\Delta T/2)$ are considered; this affects the possible entry and exit scatterers and limits the heterogeneous Green's function to a domain that depends on the entry and exit scatterer pairs.

Next, we investigate the correlation function $C_m(\Delta u)$ along the antidiagonals of $\Kmult$ as defined in Eq.~\eqref{eq:correlation}, {assuming that the medium is statistically invariant under translation.}. Using Eq.~\eqref{Km} and considering that the average over transducer position is equivalent to an ensemble average, $C_m(\Delta u)$ can be expressed as follows:

\begin{equation}
\begin{split}
\label{Cm}
   C_m(\Delta u)= & \\ 
			k_0^4 &\left \langle \iiiint  G_0(\vec{u},\ri) G_0^*(\vec{u}-\Delta \vec{u},\mathbf{r}^\prime_1) G(\ri,\rj) \right .\\
      & G^*(\mathbf{r}^\prime_1,\mathbf{r}^\prime_2)  
      G_0(\rj,\vec{u}) G_0^*(\mathbf{r}^\prime_2,\vec{u}+\Delta \vec{u})  \\
       & \left . \vphantom{\int}  \mu(\ri) \mu^*(\mathbf{r}^\prime_1)
       \mu(\rj) \mu^*(\mathbf{r}^\prime_2) \dd\ri \dd\rj \dd \mathbf{r}^\prime_1 \dd \mathbf{r}^\prime_2 \right \rangle . 
\end{split}
\end{equation}
The full calculation of $C_m(\Delta u)$ is derived in Appendix~\ref{appB}. 
The conclusion is the following: if the first and the last reflectors of a scattering sequence are located in the same \textit{resolution cell}, the corresponding wave-field is strongly correlated along the antidiagonals of the reflection matrix, as in a single scattering situation (cf. section \ref{sec:level2b}). 
The resolution cell corresponds to the focal area that would be obtained if the array was used to focus at  point $\vec{r_i}$. In our experimental configuration, the resolution cell is an ellipse centered on $\vec{r_i}$ and oriented towards the array central element. According to diffraction theory~\cite{Born}, its typical transverse and axial dimensions are given by $\Delta x =  \lambda z/A$ and $\Delta z = 7  \lambda z^2/A^2$, respectively, $A$ being the  array width.

\subsection{\label{sec:level4_2} The confocal scattering ratio}

We take advantage of the numerical scheme to study the impact of recurrent scattering on the statistical properties of the reflection matrix. The diffraction-limited recurrent contribution $\Krec$ is extracted from the total scattering matrix by selecting the paths whose first and last scatterer, $\vec{r_1}$ and $\vec{r_2}$ have overlapping resolution cells (see Fig.~\ref{fig:milieu}). 
As expected, $\Krec$ displays the long-range memory effect (see Fig.~\ref{fig:matrices}(g)), similarly to $\Kf$. In the following, we will refer to $\Kc=\Kf+\Krec$ as the \textit{confocal matrix} since it corresponds to scattering paths that start and end within the same resolution cell.
$\Kc$ is particularly relevant in view of imaging applications based on focused beamforming at emission and reception. An example of $\Kc$ is displayed in Fig.~\ref{fig:matrices}(h).

Similarly, one can define a ``conventional'' (i.e., non-recurrent) multiple scattering matrix, $\Kpmult=\Kmult-\Krec$, that  only contains the multiple scattering paths whose first and last scattering events do not occur in the same resolution cell. One example of the simulated matrix $\Kpmult$ is shown Fig.~\ref{fig:matrices}(i). It displays a random feature with short-range correlations along its antidiagonals.

In the previous section, it was shown that various orders of multiple scattering where correlated with each other; this resulted in a single-scattering ratio which could be larger than 1. We reconsider this issue: $\Kpmult$ is written as a sum of different scattering orders $\Kpn$ obtained by removing the contribution of recurrent scattering paths from each of them. The following scalar product is calculated:

\begin{equation}
\Pi_c \left (\Kpn \right ) = \frac{\left |\left\langle \Kpn | \Kc \right\rangle \right |}{\norme{\Kpn}\norme{\Kc}}.
\label{eq:pir}
\end{equation}
Fig.~\ref{fig:ps} shows the normalized scalar product $\Pi_c$ between the confocal contribution $\Kc$ and the different scattering orders $\Kpn$ of the conventional multiple scattering contribution. 
$\Pi_c$ would actually tend to 0 if the matrices where of infinite dimensions ($N \rightarrow \infty$). 
With $N=64$, the scalar product $\Pi_c$ is found to be much smaller than the initial indicator based on $\Kf$ and $\Knf$ (Eq.~\eqref{eq:pi}). This confirms that the high degree of correlation observed between $\Kf$ and $\Knf$ (simulated data) was due to recurrent scattering. As to experimental data, since the matrix filter (Eq. \eqref{ds}) is based on the memory effect, it is more relevant to interpret the reflection matrix as an addition of confocal+non-recurrent contributions rather than single+multiple scattering contributions. Then, like for the single scattering rate (Eq.~\eqref{eq:rho1}), a confocal scattering ratio can be defined as follows:
\begin{equation}
\rho_c =\frac{ \norme{\Kc}^2}{\norme{\Ktf}^2 }.
\label{eq:rhoc}
\end{equation}
The ratio $\rho_c$ is calculated with the numerical simulation and plotted in Fig.~\ref{fig:propDS}. It is consistent with the estimator $\hat{\rho}$ that can be obtained from experimental data. This highlights the fact that matrix $\Kc$ is meaningful and corresponds to the filtered matrix $\Kfilt$.

\begin{figure}[hbt!]
\centering
\includegraphics[width=8.6cm]{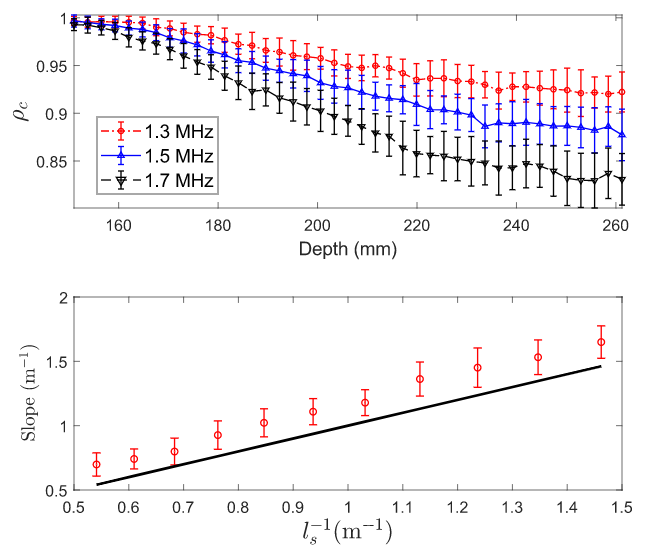}
\caption{{Confocal scattering ratio as a function of depth for three different frequencies (top). Average and standard deviation over 50 sample realizations of the slope of $\rho_{c}$ (red o) as a function of $\ell_s^{-1}$ which is also plotted in black (bottom).}}
\label{fig:pente_ls} 
\end{figure}

We observe that $\rho_c$ as well as $\hat{\rho}$ decrease linearly with depth. As shown experimentally by Fig.~\ref{fig:propDS_exp}, the slope of $\hat{\rho}(z)$ increases with frequency, which means that the multiple scattering weight appears to increase with frequency. To investigate this effect numerically, we compute the scattering mean-free path $\ell_s(\omega)$ at several frequencies in the array bandwidth (using Eq.~\eqref{eq:ls}). We then compute the average slope of $\rho_c(z)$ for each frequency. Figure \ref{fig:pente_ls} displays this slope as a function of $1/\ell_s$.
{We find that $\rho_c$ roughly scales as $-z/\ell_s$, which suggests that the scattering mean-free path is qualitatively linked to the decay of the confocal scattering ratio in the weak scattering regime considered in this work}. However, this observation is not yet supported by a theoretical demonstration and would require further investigation.

{Finally, note that the confocal scattering ratio only weakly depends on the finite width of the medium. Indeed, because of time gating, a major part of the recurrent scattering contribution corresponds to paths exploring a very few number of resolution cells. Therefore, the recurrent scattering component strongly depends on the dimensions of the resolution cells but not on the lateral size of the medium. This property also holds for the conventional multiple scattering contribution provided that the diffuse halo extension is smaller than the medium width.}

\section{\label{sec:level4_3} Link with the reflection matrix in the focused basis}

In recent studies \cite{lambertReflectionMatrixApproach2020,velichkoQuantificationEffectMultiple2020}, 
the reflection matrix has been investigated in a focused basis in order to get a local information about the scattering properties of the medium.

In this section, we use our numerical simulation to: (\textit{i}) highlight the properties of this focused reflection matrix; (\textit{ii}) build a new confocal scattering estimator in this basis; (\textit{iii}) compare the confocal scattering weights in the transducer and focused bases. 

\subsection{Reflection matrices in the focused basis}

The focused reflection matrix $\Rrr$ is obtained by a beamforming applied in emission and in reception to a set of $N_i$ points located at  depth $z$ in the scattering medium. In practice, this operation
can be experimentally achieved by applying appropriate time delays to the probe elements~\cite{lambert_ieee2}. This focusing can also be obtained by a linear projection
of the original reflection matrix $\mathbf{K}$ in the frequency domain~\cite{lambertReflectionMatrixApproach2020}:
\begin{equation}
\label{project}
    \Rrr = \Hpp^* \times \Ktf \times \Hpp^\dagger,
\end{equation}
where the symbol $\dagger$ stands for transpose conjugate. \mbox{$\Hpp = [G_0^{\prime \prime}(\vec{r},\ui,\omega)]$} is the $N_i \times N$ Green’s matrix between the array elements $\ui$ and the focal points $\mathbf{r}=(x,z)$. Each element $R(\rin,\rout,\omega)$ of $\Rrr$ is the signal that would be recorded by a virtual transducer located at $\rout=(x_\text{out},z)$ for a virtual source located at $\rin=(x_\text{in},z)$. A ballistic time gating is then performed by integrating $\Rrr$ over the signal bandwidth $\Delta \omega$~{\cite{lambertReflectionMatrixApproach2020}}:
\begin{equation}
\label{time_gating}
    \Rz(\rin,\rout) = \int_{\Delta \omega} \Rrr(\rin,\rout,\omega) \dd \omega.
\end{equation}

The diagonal elements of each matrix $\Rz(z)$, which obey
$\rin=\rout$, correspond to the confocal image that would be obtained at the corresponding depth $z$. It was also shown that the coefficients away from the diagonal carry information about aberrations \cite{lambertDistortionMatrixApproach2020} or multiple scattering \cite{lambertReflectionMatrixApproach2020}. In particular, as the information contained in the matrix $\Rrr$ is local, 2D maps of aberration or multiple scattering estimators have been proposed in biological tissues as well as in
metallic media \cite{lambertReflectionMatrixApproach2020,velichkoQuantificationEffectMultiple2020}.

\begin{figure}[hbt!]
	\centering
	\includegraphics[width=8.6cm]{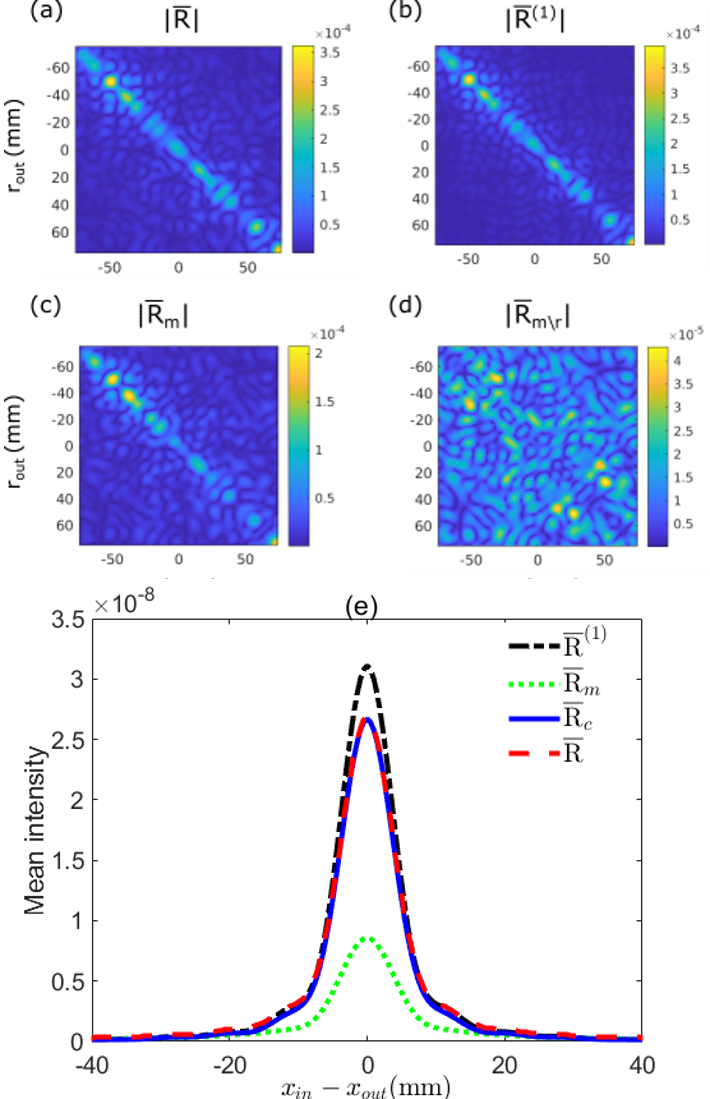}
	\caption{Absolute value of focused reflection matrices at depth $z$ = 242 mm: (a) total scattering matrix $\Rz$, (b) single scattering matrix $\Rf$, (c) multiple scattering matrix $\Rmult$, (d) non recurrent scattering matrix $\Rpmult$. (e) Mean back-scattered intensity $I$ calculated for the matrices $\Rf$, $\Rmult$,$\Rc$ and $\Rz$ following Eq.~\eqref{IrDr_fromRrDr} .}
	\label{fig:matricesFoc}
	\end{figure}

By linearity of the matrix product, the different matrices introduced in the canonical basis $\Ktf$, $\Kf$, $\Kc$, $\Kmult$, $\Kpmult$ can also be represented in the focused basis and are denoted $\Rz$, $\Rf$, $\Rc$, $\Rmult$ and $\Rpmult$ respectively. The total, single and multiple scattering focused matrices are displayed in Fig.~\ref{fig:matricesFoc} at depth $z=242$ mm. As expected, the single scattering contribution mainly emerges along the 
diagonal of $\Rz^{(1)}$ (Fig.~\ref{fig:matricesFoc}(b)). A significant part of the multiple scattering component is also found along the diagonal of $\Rz_m$ (Fig.~\ref{fig:matricesFoc}(c)). Based on our previous observation, this result is actually not so surprising and can be accounted by the predominance of recurrent scattering paths whose contribution emerges along the diagonal of $\Rz_m$.

To be more quantitative and investigate the relative part of single, recurrent and multiple scattering in the back-scattered wave-field, the mean intensity along each antidiagonal of $\Rz$ can be computed as : 
\begin{equation}
\label{IrDr_fromRrDr}
I(\Delta x )=\left \langle \left | \bar{R}(x +\Delta x /2,x-\Delta x /2,z) \right |^2 \right \rangle ,
\end{equation}
where $\langle\cdots\rangle$ denotes an average over the pairs of points $\rin=(x_\text{in},z)$ and $\rout=(x_\text{out},z)$ which are separated by the same distance $\Delta x=|x_\text{out}-x_\text{in}|$. This intensity can be calculated for each focused matrix. The corresponding intensity profiles $I(\Delta x )$ are shown in Fig.~\ref{fig:matricesFoc}(e). The value of $I_m$ at $\Delta x=0$ yields the recurrent scattering intensity. The latter quantity is thus about one fourth of the single scattering intensity at the corresponding depth. As observed previously in the transducer basis, the single and recurrent scattering intensities are not additive: the total intensity is not equal to their sum along the diagonal.

By removing recurrent scattering paths from the multiple scattering matrix $\Rz_m$, the resulting non-recurrent scattering matrix $\Rz_{m \backslash r}$ displays a random feature without any intensity peak along its diagonal (cf. Fig.~\ref{fig:matricesFoc}(d)). As before, the single and recurrent scattering paths can be rearranged in a so-called confocal matrix $\Rz_{c}$. The corresponding confocal intensity is shown to be very close to the overall intensity $I(\Delta x=0)$ along the diagonal of $\Rz$ (Fig.~\ref{fig:matricesFoc}(e)).

\subsection{Confocal scattering estimator in the focused basis}

\begin{figure}[hbt!]
	\centering
	\includegraphics[width=8.6cm]{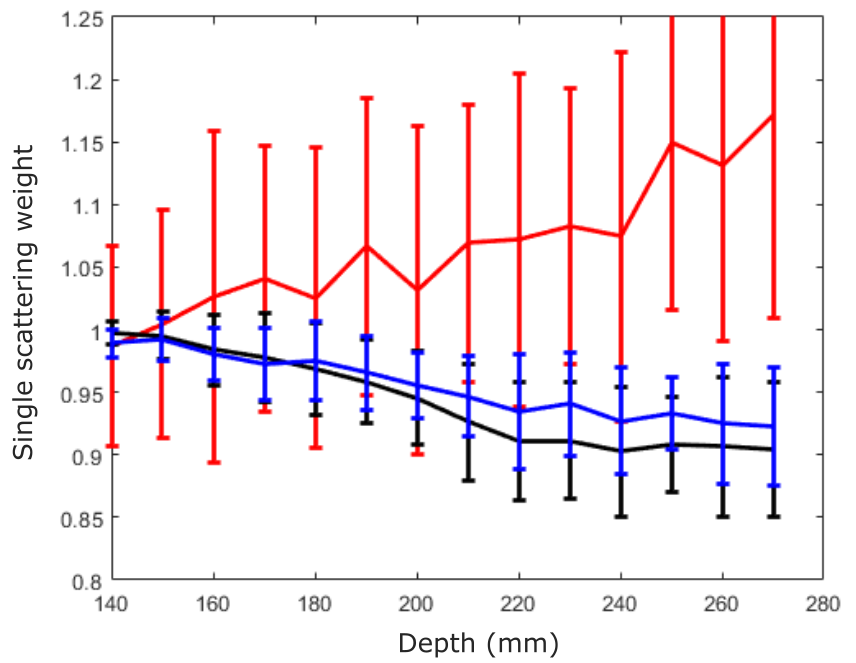}
	\caption{Average and standard deviation over 50 medium realizations of the single scattering weights $\rho_{\it f}$ (red (upper) line), $\rho_{\it fc}$ (black (bottom) line) and $\hat{\rho}_{\it f}$  (blue (intermediate) line) defined by Eqs.~\eqref{eq:rhof1}, \eqref{eq:rhofc} and \eqref{eq:rhoff}, respectively.}
	\label{fig:propDSfoc}
\end{figure}

The single and confocal scattering rates, defined by Eqs.~\eqref{eq:rho1} and \eqref{eq:rhoc} in the transducer basis, can also be built in the focused basis:
\begin{equation}
\rho_{\it f}= \frac{ \norme{\Rf}^2}{ \norme{\Rz}^2},
\label{eq:rhof1}
\end{equation}
\begin{equation}
\rho_{\it fc}= \frac{ \norme{\Rc}^2}{\norme{\Rz}^2}.
\label{eq:rhofc}
\end{equation}
The evolution of $\rho_{\it f}$ and $\rho_{\it fc}$ as a function of depth is displayed in Fig.~\ref{fig:propDSfoc} for the numerical simulation described above. Because the beamforming operation of Eq.~\eqref{project} is nearly unitary,
the values of $\rho_{\it f}$ and $\rho_{\it fc}$ are close to their counterparts, $\rho$ and $\rho_{c}$, in the transducer basis  (Fig.~\ref{fig:propDS}).

To find an estimator of these parameters, a single scattering filter has to be defined in the focused basis. Previous works 
considered a confocal gaussian filter applied to matrix $\Rz$ \cite{badonSmartOpticalCoherence2016,blondelMatrixApproachSeismic2018} in order to eliminate multiply-scattered echoes emerging far from the main diagonal. However, the shape of this filter has not been precisely linked to the intensity distribution of the single scattering matrix $\Rf$.

\begin{figure}[hbt!]

\includegraphics[width=8.6cm]{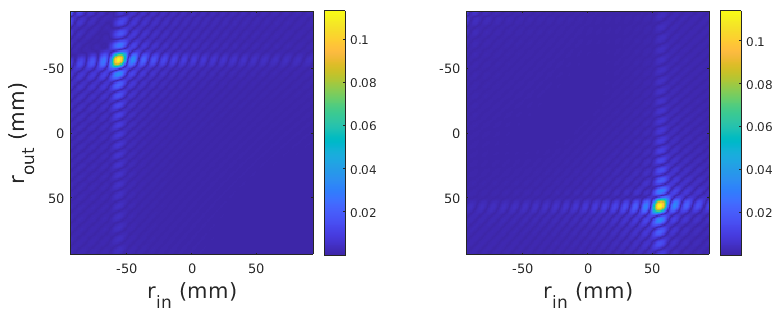}
\caption{Two elements of $\left\{ \Rll \right\}$ calculated at the simulation central frequency for two scatterers located at $z =$ 242 mm and lateral positions $x =$ -58 mm (left)  and $x =$ 54 mm (right).} 
\label{fig:chap3_matricesGénératricesBaseFocOrthonormee}
\end{figure}

To be more quantitative, an accurate single scattering space should be built in the focused basis, as previously done in the transducer basis (Eq.~\eqref{ds}). 
In the focused basis, an element of the single scattering space is the  scattering matrix for a unique scatterer located at a position $(x_l,z)$, noted $\Rl$. To span the whole  single scattering space, $N_l$ points $x_l$ have to be spread along the medium width with at least two points per unit cell. An adequate number of points must be used to avoid oversampling. The set of matrices $\Rl$ is orthogonalized using a Gram-Schmidt process in order to get an  orthonormal basis of the single scattering space that is denoted $\Rll$. Two examples of single scattering space matrices are shown in Fig.~\ref{fig:chap3_matricesGénératricesBaseFocOrthonormee}. 

The filtered matrix is then obtained by a projection of the focused total scattering matrix onto the set of matrices $\left\{ \Rll \right\}_{1 \leq l \leq N_l}$ at each frequency:
\begin{equation}
	\Rfiltf = \sum_{l=1}^{N_l} \left\langle \Rll | \Rrr \right\rangle \Rll.
	\label{eq:projfoc}
\end{equation}
$\Rfiltf$ is then integrated over a bandwidth $\Delta \omega$ to give $\Rfilt$ (Eq.~\eqref{time_gating}). An experimentally available single scattering weight estimator in the focused basis writes:
\begin{equation}
\hat{\rho}_{\it f}= \frac{ \norme{\Rfilt}^2}{ \norme{\Rz}^2}.
\label{eq:rhoff}
\end{equation}
$\hat{\rho}_{\it f}$ is compared to $\rho_{\it f}$ (Eq.~\eqref{eq:rhof1}) and $\rho_{\it fc}$ (Eq.~\eqref{eq:rhofc}) in Fig.~\ref{fig:propDSfoc}. As in the transducer basis, $\hat{\rho}_{\it f}$ is found to be a satisfying estimator of the confocal scattering ratio ${\rho}_{\it fc}$.

Interestingly, the focused basis appears more flexible compared to the canonical basis. Indeed, the time-gating operation of Eq.~\eqref{time_gating} enables an optimal selection of singly-scattered echoes associated with reflectors located at a given depth $z$. It is thus more adapted for imaging purposes than the abrupt time window originally applied  to impulse responses in the transducer basis.  

\begin{figure}[hbt!]
	\centering
	
	\includegraphics[width=8.6cm]{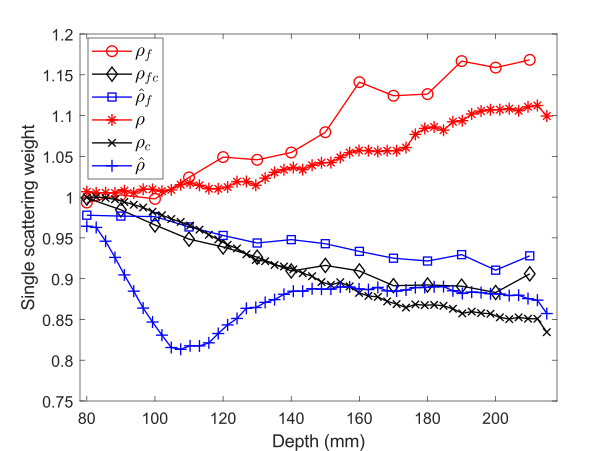}
	
	\caption{Numerical simulation beyond the paraxial approximation ($z_0=80$ mm). Average over 50 realizations of {the single scattering ratio $\rho$, the confocal scattering ratio $\rho_c$ and the estimator $\hat{\rho}$ in the canonical basis, and the corresponding values $\rho_{\it f}$, $\rho_{\it fc}$ and $\hat{\rho}_{\it f}$ in the focused basis.}}
	\label{fig:8cmFromProbe}
\end{figure}

Furthermore, the estimator $\hat{\rho}_{f}$ does not require any paraxial approximation. 
To illustrate this superiority of the focused basis, a numerical simulation is performed with a smaller distance, $z_0=80$ mm, between the probe and the front face of the sample. The single and confocal scattering weights and their estimators calculated in both bases are displayed on Fig.~\ref{fig:8cmFromProbe}. At the front face of the sample, the paraxial approximation is not valid. $\hat{\rho}$ is then not a good estimator of the confocal scattering weight $\rho_c$ since the long-range correlation along the anti-diagonals of $\Kf$ is not verified (Eq.~\eqref{eq:coherence}). Beyond the center of the medium ($z>150$ mm), the far-field approximation becomes valid and the parameters $\hat{\rho}$ and $\rho_c$ tend towards each other. On the contrary, the estimator $\hat{\rho}_{\it f}$ in the focused basis is consistent with $\rho_{\it fc}$ over the whole range of depths. This numerical simulation thus illustrates the robustness of the estimator $\hat{\rho}_{\it f}$ defined in the focused basis. 

\section{\label{sec:level6}Conclusion}

This paper conveys an improved understanding on the “single scattering rate" of backscattered waves captured by a finite-size array with elements acting both as emitters and receivers. A numerical calculation of the acoustic response of a random medium made of isotropic fluid scatterers is performed. Each term of the Born series is calculated, as well as the sum of the series providing the full reflection matrix in the frequency domain. By means of an inverse Fourier transform, the impulse responses for each scattering order can be obtained, which would not be feasible experimentally. Simple as it is, the numerical scheme sheds new light on the relative importance of single and multiple scattering contributions in the total field. Particularly, the existence of correlations between the elements of the Born series is made clear, and interpreted as a result of recurrent scattering combined with finite axial and lateral resolutions. Accordingly, the single scattering weight $\rho$ as defined by the Born series, is shown to differ from $\rho_c$, the confocal scattering weight that includes both single and recurrent scattering.
Unlike $\rho$, $\rho_c$ can be estimated from experimental data. In addition to numerical computations, experiments were carried out with ultrasound waves around 1.5 MHz in a weakly scattering forest of nylon rods. Interestingly, our results indicate that the decay of $\rho_c$ with depth could be used as a characterization tool, giving access to the scattering mean-free path $\ell_s$ in a backscattering configuration. Besides, this measurement is independent of the intrinsic absorption, as long as the duration of the time-windows $\Delta T$ is smaller than the absorption time. 

In this paper, for simplicity the study has been restricted to two-dimensional weakly scattering media for which the Born series is convergent. In order to broaden the analysis to more complex media, the numerical computation can be extended to 3D; stronger scattering regimes could also be investigated, if necessary by considering absorption in order to prevent the Born series from diverging. Although an acoustic, thus scalar, formalism is considered in this paper, the underlying physical argument is identical for other types of waves e.g, elastic or electromagnetic waves; the Born decomposition can also be applied, though in a less tractable manner, to vector waves by considering wave polarization and Green's tensors.
The issue of how much single scattering is present is essential for imaging with waves, whatever their nature: an imaging device has a finite spatial resolution, hence an elementary voxel whose dimensions depend on wavelength, depth and aperture. From the receiver's point of view, two scatterers within the same voxel are unresolved hence behave as one single super-scatterer. As a consequence, reflectivity maps obtained with an imaging device do not result only from single scattering echoes (in the sense of the Born series) but also from recurrent scattering events.  Whatever the kind of wave and the kind of media, we hope this paper brings some new insight in that respect. 

\section{\label{sec:level7}Acknowledgments}
The authors wish to thank R. Pierrat for fruitful discussions and advice, and W. Lambert for his help on the matrix analysis algorithms.  The authors are grateful for funding provided by LABEX WIFI (Laboratory of Excellence within the French Program Investments for the Future, ANR-10-LABX24 and ANR-10-IDEX-0001-02 PSL*). C.B. acknowledges financial support from Safran. A.A. acknowledges financial support from the European Research Council (ERC) under the European Union’s Horizon 2020 research and innovation programme (grant agreement No. 819261).

%

\appendix
\section{Derivation of the T-matrix for a small compressibility scatterer}
\label{appendix:tmatrix}
The incident field $\psi_0(\vec{r})$ and the total field $\psi(\vec{r})$ are linked through the Lippmann-Schwinger equation   (Eq.~\eqref{lipp} in the main text). Omitting the frequency dependence, it is written as follows :
\begin{equation}
\psi(\vec{r})= \psi_0(\vec{r}) + k_0^2 \int  G_0(\vec{r},\vec{r'}) \mu(\vec{r'}) \psi(\vec{r'}) \dd\vec{r'}.
\label{lipp_annexe}
\end{equation}
In the case of a sub-wavelength scatterer of volume $\mathcal{V}$ centered  at position $\vec{r_s}$, the field $\psi(\vec{r})$ can be considered as constant inside the particle, so that Eq.~\eqref{lipp_annexe} is simplified as :
\begin{equation}
\psi(\vec{r})= \psi_0(\vec{r}) + k_0^2 
\mu(\vec{r_s}) \psi(\vec{r_s}) \int_\mathcal{V} G_0 (\vec{r},\vec{r'}) \dd\vec{r'}.
\label{lipp_annexe_small}
\end{equation}

By definition, the $\mathbf{T}$-matrix 
links the incident and scattered fields through the following equation:
\begin{equation}
  \psi(\vec{r})= \psi_0(\vec{r}) 
   + \iint G_0 (\vec{r},\vec{r_1}) T(\vec{r_1},\vec{r_2}) 
  \psi_0(\vec{r_2}) \dd\vec{r_1} \dd\vec{r_2}  .
\end{equation}
For the sub-wavelength scatterer at position $\vec{r_s}$ :
\begin{equation}
T_s(\vec{r},\vec{r^\prime}) = \T \delta(\vec{r}-\vec{r_s}) \delta(\vec{r^\prime}-\vec{r_s}).
\end{equation}
It follows that 
\begin{equation}
\psi (\vec{r}) = \psi_0(\vec{r}) +  \T  G_0 (\vec{r},\vec{r_s})  \psi_0(\vec{r_s}) .
\end{equation}

By replacing the field $\psi_0(\vec{r_s})$ by its expression given by \eqref{lipp_annexe_small}  taken at $\vec{r}=\vec{r_s}$   into the preceding equation it comes : 
\begin{equation}
\begin{split}
   \psi (\vec{r}) = &\psi_0(\vec{r})\\
                     + & \T G_0 (\vec{r},\vec{r_s}) \left( 1- k_0^2 \mu(\vec{r_s}) \int_\mathcal{V} G_0 (\vec{r_s},\vec{r'}) \dd\vec{r'} \right)  \psi(\vec{r_s})  .
\end{split}
\end{equation}
For a point $\vec{r}$ far from the particle, $G_0$ can be considered as constant in the scatterer volume, so that  \eqref{lipp_annexe_small} becomes:
\begin{equation}
          \psi (\vec{r}) = \psi_0 (\vec{r}) +\mathcal{V} \T G_0  (\vec{r},\vec{r_s}) \psi(\vec{r_s}).
\end{equation}
Thus, by identifying terms in the last two equations : 
\begin{equation}
          \T \left( 1- k_0^2 \mu(\vec{r_s}) \int_\mathcal{V} G_0 (\vec{r_s},\vec{r'}) \dd\vec{r'} \right) = k_0^2 \mathcal{V} \mu(\vec{r_s}).
\end{equation}

For a  scatterer of radius $a$, the integral of the Green's function in two dimensions can be approximated as

\begin{equation}
   \int_\mathcal{V} G_0 (\vec{r_s},\vec{r'}) \dd\vec{r'} \approx \frac{-\imath \pi a^2}{4}.
\end{equation}

This provides an analytical expression of the scattering coefficient $\T$ 
\begin{equation}
          \T \approx \frac {k_0^2\mu(\vec{r_s})  \pi a^2}{ 1 + \frac{\imath}{4} k_0^2 \mu(\vec{r_s})\pi a^2  }.\end{equation}

\section{\label{appB}Recurrent scattering correlation}

A multiple scattering sequence involves at least two distinct scatterers. Let us denote $\ri$ and $\rj$ their positions. Taking the individual scatterer as the unit scattering cell, we sum all over possible entry and exit scatterer pairs whose positions are compatible with the time-gating condition. Using the properties of the $\mathbf{T}$-matrices, the $(i,j)$ element of $\Kmult$ may be written as: 
\begin{equation}
\begin{split}
\label{Km2}
      K_m(u_i,u_j) =   k_0^2 \iint\limits G_0(\ui,\ri) & \mu(\ri) G(\ri,\rj) \mu(\rj)\\ &G_0(\rj,\uj) \dd\ri\dd\rj,
\end{split}
\end{equation}

Only the multiple scattering sequences whose path lengths are comprised between $c_0(T-\Delta T/2)$ and $c_0(T+\Delta T/2)$ are considered; this affects the possible entry and exit scatterers and limits the heterogeneous Green's function to a domain that depends on the entry and exit scatterer pairs. Next, we investigate the correlation function $C_m(\Delta u)$ defined in Eq.~\eqref{eq:correlation}. Using Eq.~\eqref{Km2}, $C_m(\Delta u)$ can be expressed as follows: 
\begin{widetext}
\begin{equation}
\begin{split}
\label{Cm2}
      C_m(\Delta u)=  k_0^4 
    &\left  \langle \iiiint  G_0(\vec{u},\ri)
    G_0^*(\vec{u}-\Delta\vec{u},\vec{r'_1}) G(\ri,\rj)  
     G^*(\vec{r'_1},\vec{r'_2})G_0(\rj,\vec{u}) G_0^*(\vec{r'_2},\vec{u}+\Delta \vec{u})  \right .\\
     &\left . \vphantom{\int} \mu(\ri) \mu^*(\vec{r'_1}) 
       \mu(\rj) \mu^*(\vec{r'_2}) \dd\ri\dd\rj \dd \vec{r'_1} \dd\vec{r'_2}
      \right \rangle .
\end{split}
\end{equation}
\end{widetext}
Note that the medium is assumed to be statistically invariant by translation. Hence, the correlation coefficient $C_m$ does not depend on the position $\vec{u}$.

To go further, we assume that the scatterers positions are independent random variables. In the weak scattering regime ($k\ell_s \gg 1$), most contributions to the correlation function, $\langle G(\ri,\rj) G^*(\vec{r'_1},\vec{r'_2}) \rangle$, will cancel out in the above ensemble average. The only contributions to survive this average are those for which the wave and its complex conjugate experience identical paths. This condition is achieved if the wave and the complex conjugate visit the same scatterers either in the same order (ladder diagrams), or in reversed order (maximally crossed diagrams). The correlation function can thus be decomposed as the sum of two terms~\cite{akkermansMesoscopicPhysicsElectrons2007}:

\begin{equation}
\begin{split}
&\left\langle G(\ri,\rj)G^\ast(\ri^\prime,\rj^\prime)\right\rangle \propto
k_0^{-4} P(\ri,\rj) \\  &\left [ \delta(\ri-\ri^\prime)\delta(\rj-\rj^\prime)+\delta(\ri-\rj^\prime)\delta(\rj-\ri^\prime)\right],\label{eq:correldirac}
\end{split} 
\end{equation}
where $P(\ri,\rj) $ is the mean intensity Green's function. In a statistically homogeneous medium, this quantity only depends on $|\ri-\rj|$ and on the time lapse between scattering events at $\ri$ and $\rj$. The left and right terms in Eq.~\eqref{eq:correldirac} correspond to the contribution of identical and reciprocal scattering paths, respectively. {Ladder and crossed diagrams contribute equally to $C_m(\Delta u)$ (Eq.~\ref{Cm2}) whose expression can be simplified as follows:
\begin{widetext}
\begin{equation}
      C_m(\Delta u) = 2 \langle |\mu|^2 \rangle ^2 \iint G_0(\vec{u},\ri) G_0^*(\vec{u}-\Delta \vec{u},\ri) P(|\ri-\rj|) 
      G_0(\rj,\vec{u}) G_0^*(\rj,\vec{u}+\Delta \vec{u}) \dd\ri \dd\rj ,
      \label{eq:chap2_cm}
\end{equation}
\end{widetext}}

Considering the Green's function in free space in 3D or in 2D far-field (Eq.~\eqref{eq:chap2_solutionGreen}), the argument $\phi$ in the integrand of $C_m$ (Eq.~\eqref{eq:chap2_cm}) writes: 
\begin{equation}
\begin{split}
      {\phi} = & k_0 \left( |\vec{u}-\ri| - |\vec{u}-\Delta \vec{u} - \ri| \right. \\ & \left . + |\vec{u}-\rj| - |\vec{u}+\Delta \vec{u} - \rj| \right).
      \end{split}
\end{equation}
In the far field, a first-order expansion gives:
\begin{equation}
\label{phiL}
      {\phi} \approx k_0 \left [  \Delta u \left( \frac{u-x_1}{z_1}-\frac{u-x_2}{z_2} \right) \right].
\end{equation}
Let us define $z_0=(z_2+z_1)/2$ and  $dz =(z_2 - z_1)/2$. Using a first-order approximation in $dz/z_0$ we obtain:
\begin{equation}
\label{phiL2}
    \begin{split}
      {\phi} \approx k_0 \left  \lbrace  \frac{\Delta u}{z_0} \left[ x_2-x_1 + \frac{dz}{z_0}(2u-x_1-x_2) \right ] \right \rbrace.
    \end{split}
\end{equation}
It leads to the following expression for $C_m$ in 2D: 
\begin{equation}
\begin{split}
  & C_m(\Delta u) =  \frac{2 \langle |\mu|^2  \rangle ^2  }{(8 \pi k_0 z_0)^2}  
      \iint P(|\ri-\rj|)  \\ & \exp \left \lbrace \imath \frac{k_0 \Delta u}{z_0}\left [ (x_2-x_1)+(2u-x_1-x_2)\frac{dz}{z_0}\right] \right \rbrace \dd\ri\dd\rj .
      \label{eq:chap2_correlationRecurrente}
\end{split}
\end{equation}
Scattering paths such that the argument $\phi$ is small compared to $\pi$ will result in a correlation coefficient $C_m$ independent of $\Delta u$ at the scale of the array. In other words, such scattering paths will induce a long-range correlation of the reflected field along the anti-diagonals of $\mathbf{K}$. A sufficient condition for this to occur is that the transverse distance between the first and last
scatterers $|x_1 -x_2|$ is smaller than the transverse size of the resolution cell $\Delta x \sim \lambda z_0/A$, with
$A$ the aperture of the probe, and that the axial distance $|z_1 -z_2|$ between them is less than the
depth of field $\Delta z \sim  7 \lambda z_0^2/A^2$. In other words, these are the scattering paths whose first and last scatterers are contained in the same resolution cell, that is to say the so-called recurrent scattering paths. If the mean intensity Green's function $P(|\ri-\rj|)$ has a characteristic support of transverse extension $\Delta x_P < \Delta x $ and axial dimension $\Delta z_P < \Delta z$, these recurrent scattering paths
predominate and the reflection matrix shows a long-range correlation along its anti-diagonals, analogous to that obtained in the single scattering regime. This effect has already been observed in strongly scattering media near the Anderson transition~\cite{aubryRecurrentScatteringMemory2014}.

\section{\label{appC}{Coherent back-scattering peak}}
{Using Eqs.~\ref{Km2} and \ref{eq:correldirac}, the mean backscattered intensity, $I(\Delta u)= \langle | K_m(\vec{u},\vec{u}+\Delta \vec{u})|^2 \rangle $, can be decomposed as the sum of a ladder ($I_L$) and a crossed ($I_C$) integral:
\begin{equation}
    I(\Delta u)=I_L(\Delta u)+I_C(\Delta u)
\end{equation}
with $I_L$ corresponding to the incoherent summation of multiple-scattering paths intensity, such that
\begin{widetext}
\begin{equation}
      I_{L}(\Delta u) =  \langle |\mu|^2 \rangle ^2 \iint |G_0(\vec{u},\ri)|^2 P(|\ri-\rj|) 
         |G_0(\rj,\vec{u}+\Delta \vec{u}) |^2\dd\ri\dd\rj ,
      \label{eq:diff}
\end{equation}
and $I_C$ resulting from the interference between multiple-scattering paths and their reciprocal counterparts
\begin{equation}
   I_{C}(\Delta u) =  \langle |\mu|^2 \rangle ^2 \iint G_0(\vec{u},\ri) G_0^*(\vec{u},\rj) P(|\ri-\rj|) 
       G_0(\rj,\vec{u}+\Delta \vec{u}) G_0^*(\ri,\vec{u}+\Delta \vec{u}) \dd\ri\dd\rj .
      \label{eq:cbs}
\end{equation}
\end{widetext}
The latter term accounts for the coherent back-scattering peak~\cite{Tourin1997,aubryCoherentBackscatteringFarfield2007}. Considering the Green's function in free space in 3D or in 2D far-field (Eq.~\eqref{eq:chap2_solutionGreen}), the argument $\phi_{C}$ in the integrand of $I_C$ (Eq.~\eqref{eq:cbs}) writes: 
\begin{equation}
\begin{split}
      \phi_{C} = & k_0 \left( |\vec{u}-\ri| - |\vec{u}+\Delta \vec{u} - \ri| \right. \\ & \left . - |\vec{u}-\rj| + |\vec{u}+\Delta \vec{u} - \rj| \right).
      \end{split}
\end{equation}
In the far field, a first-order expansion gives:
\begin{equation}
      \phi_{C} \approx - k_0 \left [  \Delta u \left( \frac{u-x_1}{z_1}-\frac{u-x_2}{z_2} \right) \right].
\end{equation}
This expression is strictly the opposite of Eq.~\ref{phiL}, hence $ \phi_{C} \simeq - \phi$ in the far-field. Using Eq.~\ref{phiL2}, a final expression can be found for $I_{C}$ in 2D:
\begin{equation}
\begin{split}
  &I_{C}(\Delta u) =   \frac{\langle |\mu|^2  \rangle ^2 }{(8 \pi k_0 z_0)^2}  
      \iint  P(|\ri-\rj|) \\ &  \exp \left \lbrace - \imath \frac{k_0 \Delta u}{z_0}\left [ (x_2-x_1)+(2u-x_1-x_2)\frac{dz}{z_0}\right] \right \rbrace \dd\ri\dd\rj .
      \label{eq:Ic}
\end{split}
\end{equation}
By comparing the latter expression with Eq.~\ref{eq:chap2_correlationRecurrente}, we find that the correlation coefficient $C_m$ along the antiagonals of $\Kmult$ is similar to that of the coherent back-scattering peak $I_C$: $I_{C}(\Delta u) \propto C_m(\Delta u) $. Both quantities thus exhibit the same dependence in $\Delta u$ in the far-field.}

\end{document}